\documentclass[aps,prc,twocolumn,floatfix,superscriptaddress,nofootinbib]{revtex4-2}

\usepackage{graphicx}
\usepackage[urlcolor=green]{hyperref}
\usepackage{amsmath}
\usepackage{physics}
\usepackage{xcolor}
\usepackage{placeins}

\newcommand{\Nsrc}{N_{\rm src}}
\newcommand{\Nev}{N_\mathrm{ev}}
\newcommand{\ssrc}{\sigma_\mathrm{src}}

\usepackage[maxfloats=256]{morefloats}
\begin{document}

\author{Nicolas Borghini} \email{borghini@physik.uni-bielefeld.de}
\affiliation{Fakult\"at f\"ur Physik, Universit\"at Bielefeld, D-33615 Bielefeld, Germany}
\author{Hendrik Roch} \email{roch@fias.uni-frankfurt.de}
\affiliation{Frankfurt Institute for Advanced Studies, Ruth-Moufang-Strasse 1, 60438 Frankfurt am Main, Germany}
\affiliation{Department of Physics and Astronomy, Wayne State University, Detroit, Michigan 48201, USA}
\author{Alicia Sch\"utte} \email{alicia.schuette@uni-bielefeld.de}
\affiliation{Fakult\"at f\"ur Physik, Universit\"at Bielefeld, D-33615 Bielefeld, Germany}

\title{Statistical analysis of the fluctuations of an initial-state model with independently distributed hot spots}

\begin{abstract}
We determine the uncorrelated modes that characterize the fluctuations in a semi-realistic model for the initial state of high-energy nuclear collisions, consisting of hot spots whose positions are distributed independently. 
Varying the number of hot spots, their size, and the weights with which they contribute to the initial state, we find that the parameter that has the largest influence on the relative importance of the fluctuation modes is the source size, with more extended hot spots leading to a more marked predominance of the principal modes.
\end{abstract}

\maketitle

\section{Introduction}
\label{s:intro}

When two nuclei collide at high energy, they are strongly Lorentz contracted, so that each of them only sees an instantaneous snapshot of the wave function of the other. 
This results in event-by-event fluctuations in the positions of the colliding degrees of freedom, which were shown to lead to long-range azimuthal correlations between the emitted particles in the final state of the dynamical evolution of the created system~\cite{Miller:2003kd,Andrade:2006yh,PHOBOS:2006dbo,Alver:2010gr}.

A large variety of approaches have been proposed to describe the energy- or entropy-density profile, with the possible addition of conserved charges, resulting from the collision of two nuclei shortly after these have passed through each other, and that may be used as ``initial state'' for a subsequent dynamical evolution~\cite{Socolowski:2004hw,Drescher:2006ca,Miller:2007ri,Broniowski:2007nz,Broniowski:2007ft,Bratkovskaya:2011wp,Schenke:2012wb,Schenke:2012hg,Paatelainen:2012at,Albacete:2014fwa,Moreland:2014oya,Loizides:2014vua,SMASH:2016zqf,Giacalone:2019kgg,Shen:2020jwv,Schafer:2021csj,Shen:2022oyg,Garcia-Montero:2023gex}. 
The models extend from physics-agnostic parametrizations to effective theories of quantum chromodynamics (QCD) at high energy, relying either on nucleon or on subnucleonic degrees of freedom, and they may involve some dynamical evolution --- especially at lower collision energies --- or be purely static, resulting in either two- or more recently three-dimensional configurations.

Of paramount importance is the connection between initial-state properties and final-state observables, mostly in the form of multiparticle correlations~\cite{Luzum:2013yya,Giacalone:2023hwk}: 
A systematic analysis of the latter may lead to a satisfactory enough determination of the initial state to pinpoint the dynamic properties of the strongly-interacting created system. 
In that spirit, several characterizations of the initial geometry have been proposed, either using some a priori expansion~\cite{Coleman-Smith:2012kbb,Floerchinger:2014fta,Mazeliauskas:2015vea,Floerchinger:2018pje}, or relying on a basis of ``uncorrelated fluctuation modes'' depending on the set of events under consideration~\cite{Borghini:2022iym}. 

In the present study, we use a toy model mimicking several features predicted for the initial state in realistic models, and we investigate how varying these features affects the fluctuation modes advocated in Ref.~\cite{Borghini:2022iym}. 
For that purpose, the principle of the decomposition is first recalled in Sec.~\ref{s:mode-decomposition}.  
We then introduce in Sec.~\ref{s:IHSM} the simple initial-state model that we use and discuss its numerical implementation.
The details of our simulations together with their results are presented in Sec.~\ref{s:results} and in Appendix~\ref{s:c_l}. 
We also include a comparison to results from the Monte Carlo (MC) Glauber model of Ref.~\cite{Borghini:2022iym}.
Eventually, the main findings are summarized and discussed in Sec.~\ref{s:Discussion}.

\section{Initial state fluctuations and their mode decomposition}
\label{s:mode-decomposition}

Within a model of the initial state, one can generate an ensemble of $\Nev$ configurations $\{\Phi^{(i)}(\vb{x})\}$ --- where $\vb{x}$ denotes the position in two or three dimensions, according to the model --- with similar characteristics: for instance, initial energy-density profiles for a given collision system at the same impact parameter, or corresponding to events within a definite centrality class.
It seems rather natural to write down each configuration as the sum of an average initial state $\bar{\Psi}(\vb{x})$, defined as the arithmetic mean of the $\{\Phi^{(i)}(\vb{x})\}$, and the departure $\delta\Phi^{(i)}(\vb{x})$ from this average:
\begin{equation}
\label{eq:Phi=Psi+dPhi}
\Phi^{(i)}(\vb{x}) = \bar{\Psi}(\vb{x}) + \delta\Phi^{(i)}(\vb{x})
\end{equation}
with
\begin{equation}
\bar{\Psi}(\vb{x}) \equiv \frac{1}{\Nev} \sum_{i=1}^{\Nev} \Phi^{(i)}(\vb{x}).
\label{eq:avg_state}
\end{equation} 
The $\{\delta\Phi^{(i)}(\vb{x})\}$ are referred to as the fluctuations about the average $\bar{\Psi}(\vb{x})$.
The decomposition~\eqref{eq:Phi=Psi+dPhi} is in particular useful to investigate the influence of different types of initial-state fluctuations on observables in the final state of the dynamical evolution of the system~\cite{Floerchinger:2013rya,Floerchinger:2013vua,Floerchinger:2013hza,Floerchinger:2013tya,Floerchinger:2014fta,Mazeliauskas:2015vea,Floerchinger:2018pje,Floerchinger:2020tjp}.
For such mode-by-mode studies, the customary approach is to describe the fluctuations via their decomposition on a predetermined basis, e.g.\ via a Bessel--Fourier series, as advocated in Refs.~\cite{Coleman-Smith:2012kbb,Floerchinger:2014fta} for two-dimensional profiles.

Instead of using such an a-priori given basis, it was shown in Ref.~\cite{Borghini:2022iym} that it is possible to find a basis of ``fluctuation modes'' $\{\Psi_l(\vb{x})\}$ that is closely tied to the fluctuations $\{\delta\Phi^{(i)}(\vb{x})\}$ under study, in the following sense. 
Writing each fluctuation $\delta\Phi^{(i)}(\vb{x})$ as 
\begin{equation}
\label{eq:evt-fluct_vs_modes}
\delta\Phi^{(i)}(\textbf{x}) = \sum_l c_l^{(i)} \Psi_l(\textbf{x}),
\end{equation}
the expansion coefficients $\{c_l^{(i)}\}$ obey the conditions
\begin{equation}
\label{eq:mean_and_variance_cl}
\frac{1}{\Nev} \sum_{i=1}^{\Nev} c_l^{(i)} = 0
\qand
\frac{1}{\Nev} \sum_{i=1}^{\Nev} c_l^{(i)} c_m^{(i)} = 0
\end{equation}
for all $l$ and $l\neq m$.
That is, they appear as the realizations of uncorrelated centered random variables $\{c_l\}$. 
In turn, the fluctuation modes are such that
\begin{equation}
\int\!\Psi_l(\vb{x})_{}\Psi_m(\vb{x})\dd{\vb{x}} = 0
\end{equation}
for $l\neq m$, which may be seen as an orthogonality condition. 
In that sense, the overlap integral of a mode $\Psi_l$ with itself is its squared norm and will be denoted by $\norm{\Psi_l}^2$. 
According to Eq.~\eqref{eq:evt-fluct_vs_modes}, the contribution of mode $\Psi_l$ to the fluctuation $\delta\Phi^{(i)}$ on the $i$-th configuration is the product of the mode with the expansion coefficient $c_l^{(i)}$. 
That is, the importance of the contribution can be assigned either to the expansion coefficient or to the mode. 
In Ref.~\cite{Borghini:2022iym} it was chosen to let the modes have different norms and to work with expansion coefficients with unit variance, so that the conditions~\eqref{eq:mean_and_variance_cl}, expressed for the random variables $\{c_l\}$, become
\begin{equation}
\label{eq:mean_and_variance_cl2}
\expval{c_l} = 0
\qand
\expval{c_l c_m} = \delta_{lm}, 
\end{equation}
where $\expval{\cdots}$ denotes the statistical average over events.
That is, the basis of fluctuation modes is not orthonormal, only orthogonal, and the norm $\norm{\Psi_l}\equiv\sqrt{\lambda_l}$ of each mode is a measure of its typical contribution to the fluctuations.

The configurations $\{\Phi^{(i)}(\vb{x})\}$, and therefore the corresponding fluctuations $\{\delta\Phi^{(i)}(\vb{x})\}$, generally have a physical dimension. 
Fixing the typical size of the expansion coefficients is consistent with choosing them dimensionless, thus leaving the physical dimension in the modes and their norms.
That is, the squared norms $\{\lambda_l\}$ are actually dimensionful.
Since this means that their values depend on a choice of units, one can alternatively characterize the importance of the modes by the dimensionless ratios
\begin{equation}
w_l \equiv \frac{\sqrt{\lambda_l}}{\sum_l\sqrt{\lambda_l} + \norm{\bar{\Psi}}}, 
\label{eq:w_l}
\end{equation}
with $\norm{\bar{\Psi}}$ the norm of the average initial state~\eqref{eq:avg_state}.
For consistency, we also define
\begin{equation}
\bar{w} \equiv \frac{||\bar{\Psi}||}{\sum_l\sqrt{\lambda_l} + ||\bar{\Psi}||}, 
\label{eq:wbar}
\end{equation}
which quantifies the relative ``weight'' of the average initial state. 
Clearly, these quantities are defined such that the sum of all $w_l$ and of $\bar{w}$ equals 1.

In Ref.~\cite{Borghini:2022iym}, energy-density profiles for Pb-Pb collisions at $\sqrt{s_{_{N\hspace{-.1em}N}}}=5.02$~TeV were generated with two different models --- a MC Glauber model with energy deposition controlled by the local numbers of participants and binary collisions~\cite{dEnterria:2020dwq}, and a saturation-based approach. 
For configurations at a fixed impact parameter, it was observed that the spectra of the weights $w_l$, ordered by decreasing value, significantly differ within the two models: 
the $w_l$-spectrum is steeper for the fluctuation modes of the Glauber model, and this holds at two different impact-parameter values, see Fig.~3 of Ref.~\cite{Borghini:2022iym}. 
Physically, this means that ``higher'' fluctuation modes, defined as those with a smaller $w_l$, contribute less to initial-state fluctuations in the Glauber model than in the saturation-based one. 
Yet no attempt was made in Ref.~\cite{Borghini:2022iym} to investigate which properties of the initial profiles are reflected in the behavior of the $w_l$-spectrum, which is what we address in the present study.
For that purpose, we depart from initial-state models motivated by phenomenology, as we now discuss.

\section{Independent hot spot model}
\label{s:IHSM}

We wish to use a semi-realistic toy model with well-defined features that we may change at will---which also means that we study setups that are obviously not realized in actual collisions of nuclei. 
For simplicity, we only consider transverse initial-state profiles, so that vectors, denoted in boldface, are from now on two-dimensional.

\subsection{Description of the model}
\label{ss:model_th}

Specifically, we assume that each initial state $\Phi^{(i)}(\vb{x})$ is a superposition of $\Nsrc^{(i)}$ local sources, which we shall refer to as ``hot spots''. 
Each hot spot, labeled by a subscript $k$, is distributed with some source function $s^{(i)}_k$ about a position $\vb{x}_k^{(i)}$ in the transverse plane, and it contributes with a weight $\varpi_k^{(i)}$ to the initial profile~\cite{Blaizot:2014wba,Blaizot:2014nia}:
\begin{equation}
\Phi^{(i)}(\vb{x}) = \sum_{k=1}^{\Nsrc^{(i)}} \varpi_k^{(i)} s^{(i)}_k\big(\vb{x} - \vb{x}_k^{(i)}\big).
\label{eq:IPSM_density}
\end{equation}
In most of the scenarios we study hereafter, the source function will be the same for all hot spots in all initial-state profiles, in which case one can more compactly denote it by $s(\vb{x} - \vb{x}_k^{(i)})$, as we now do for better readability. 
Our source functions are normalized to unity---possible differences in the contributions of the hot spots are encoded in the weights $\varpi_k^{(i)}$---, and to allow straightforward calculations we take them to be Gaussian-distributed with a width $\ssrc$
\begin{equation}
\label{eq:s(x)}
s(\vb{x}-\vb{x}_k) = \frac{1}{2\pi\ssrc^2}{\rm e}^{-(\vb{x}-\vb{x}_k)^2/2\ssrc^2}.
\end{equation}
In the limit of pointlike sources, i.e.\ a vanishing $\ssrc$, the source function becomes $s(\vb{x}-\vb{x}_k) = \delta^{(2)}(\vb{x}-\vb{x}_k)$.
The source width $\ssrc$ of our toy approach corresponds to the typical size of the transverse area over which energy is deposited by individual nucleon-nucleon or parton-parton collisions in usual initial-state models for high-energy heavy-ion collisions.
In modern Bayesian analyses, this typical size is often a free parameter, whose value has a significant impact on the determination of other parameters like the transport coefficients of the created hot and dense QCD matter (for a short review, see Ref.~\cite{Giacalone:2022hnz} and references therein).

Another ingredient in Eq.~\eqref{eq:IPSM_density} is the distribution of the hot-spot positions $\{\vb{x}_k^{(i)}\}$ in an event. 
They are assumed to be drawn from a random distribution $f_1(\vb{x}_k)$, which in most scenarios investigated in this paper will be a two-dimensional Gaussian.
In a couple of cases, we shall consider different widths $\sigma_x\leq\sigma_y$ along two orthogonal directions, see Eq.~\eqref{eq:IPSM_f_1} below. 
Yet, in most cases we assume that $f_1$ is rotationally invariant, with a unique width $\sigma_x=\sigma_y$.

Since a single-variable distribution is used for the hot-spot positions, they are a priori independent, which is why we call the approach ``independent hot spot model'' (IHSM). 
This is basically the model first introduced with pointlike sources in Ref.~\cite{Bhalerao:2006tp} and further elaborated upon in Ref.~\cite{Bhalerao:2011yg} (with Gaussian sources), and in Ref.~\cite{Blaizot:2014wba} (with fluctuating number of sources and weights). 
Due to its relative simplicity, variants of the model --- also referred to as ``independent cluster model'' --- were used to compute (semi-)analytically fluctuations of (multiparticle) initial-state eccentricities~\cite{Bzdak:2013rya,Basar:2013hea,Bzdak:2013raa,Blaizot:2014nia} or correlators of participant-plane angles from different harmonics~\cite{Jia:2012ju}.
Yet in the numerical simulations described hereafter, we actually recenter the generated profiles $\Phi^{(i)}(\vb{x})$, which induces (small) correlations between the positions. 

The last element in the generation of initial profiles~\eqref{eq:IPSM_density} is the weight of each source. 
They account for the intrinsic randomness in the physical mechanism --- collisions between nucleons or subnucleonic degrees of freedom --- producing the initial state.
Such fluctuating weights are implemented in the MC Glauber underlying GLISSANDO~\cite{Broniowski:2007nz,Rybczynski:2013yba,Bozek:2019wyr} and are also included (in the thickness function of each nucleus) in the T$_{\rm R}$ENTo generator~\cite{Moreland:2014oya}.
In this paper, we assume either a constant weight, taken equal to 1 since our initial profiles are not tailored for phenomenology, or a value taken from a probability distribution $p(\varpi)$ such that the average of $\varpi$ over its range of values is 1.
Note that any dependence of the hot-spot weight on position can be absorbed in the distribution $f_1$.

With the above specifications, and assuming that the sample average of the weights $\varpi_k^{(i)}$ over the $\Nsrc^{(i)}$ sources coincides with its statistical average, the integral of the profile~\eqref{eq:IPSM_density} over the transverse plane yields the corresponding number of hot spots:
\begin{equation}
\int\!\Phi^{(i)}(\vb{x})\dd[2]{\vb{x}} = \Nsrc^{(i)}.
\label{eq:int_transverse_dens_IHSM}
\end{equation}

In turn, the average initial state~\eqref{eq:avg_state} reads
\begin{equation}
\bar{\Psi}(\vb{x}) = \expval{\Nsrc}\!\int\!s(\vb{x}-\vb{x}_1)_{}f_1(\vb{x}_1)\,\dd[2]{\vb{x}_1},
\label{eq:avg_state2a} 
\end{equation}
with $\expval{\Nsrc}$ the mean number of hot spots, and where the effect of recentering the configurations is ignored.%
\footnote{Repeating the calculations detailed in Appendix~D of Ref.~\cite{Blaizot:2014wba}, one finds that the effect of recentering is to replace $\sigma_x^2$ and $\sigma_y^2$ by their product with $1-1/\!\expval{\Nsrc}^2$. 
Since we always consider of the order of 50 hot spots or more in our simulations, the change from recentering constitutes a relative correction by a factor smaller than $10^{-3}$, which we neglect. 
Note that our Eq.~\eqref{eq:avg_state2a} is Eq.~(C2) of Ref.~\cite{Blaizot:2014wba} with (on average) unit weights.}

To conclude this section, let us summarize the parameters of the model. 
The Gaussian source function $s(\vb{x})$ is entirely characterized by its width $\ssrc$. 
The (also Gaussian) probability distribution $f_1(\vb{x})$ for the hot-spot centers is determined by $\sigma_x$ and $\sigma_y$. 
A given initial-state profile is also characterized by its number of sources $\Nsrc$. 
Eventually, the relative importance of the hot spots is given by weights $\varpi$ that may fluctuate with a probability density $p(\varpi)$.

\subsection{Numerical implementation}
\label{ss:model_num}

To determine the fluctuation modes corresponding to the IHSM with given parameters, we simulated the model numerically.
For that, we discretized the transverse plane with a spatial grid comprising $N_s^2=128\times 128$ sites, each separated by a spacing $a \simeq 0.24$~fm.\footnote{More accurately, $a = 30/(N_s-1)$~fm. The length unit is strictly speaking irrelevant, since we do not perform any phenomenology.}

To generate an initial-state profile $\Phi(\vb{x})$ --- where for brevity we suppressed the superscript $(i)$ ---, we sample iteratively the positions $\{\vb{x}_k\}$ of the hot spot centers from the corresponding probability distribution $f_1(\vb{x})$, using an acceptance-rejection algorithm in the sampling process, where the origin of coordinates lies in the middle of the grid. 
The centers of the hot spots are generally not restricted to coincide with the grid points. 
The only exception is when we simulate pointlike sources, whose specific implementation is described at the end of this section.
In Secs.~\ref{ss:simulation_setup}--\ref{ss:results_modes}, the distribution of hot-spot centers is Gaussian:
\begin{equation}
f_1(\vb{x}) \equiv \frac{1}{2\pi\sigma_x\sigma_y}\exp\left[-\frac{x^2}{2\sigma_x^2}-\frac{y^2}{2\sigma_y^2}\right].
\label{eq:IPSM_f_1}
\end{equation}
In all simulations we use $\sigma_y=4$~fm, while $\sigma_x$ is either 4~fm, yielding a rotationally invariant distribution, or 2~fm. 
Since our grid extends about 15~fm along the $x$- and $y$-directions, hot-spot positions outside the grid are an extremely rare occurrence, in which case we simply discard the position and generate a new one. 

For the hot-spot widths $\ssrc$, we shall consider several values: $\ssrc=0$ (pointlike sources, see later below), 0.3, and 0.7~fm. 
We also performed simulations with a fluctuating $\ssrc$.
Given the finite extent of the grid, we cannot simulate the whole Gaussian source function~\eqref{eq:s(x)}; 
accordingly, we truncate it at a distance of $3\ssrc$ from the center $\vb{x}_k$
The values of $s(\vb{x}-\vb{x}_k)$ at the grid sites within this range are stored and summed over.
Multiplying the resulting sum with the grid-spacing squared $a^2$ amounts to a numerical integration of the source function. 
This integral will generally not give 1, contrary to the analytical model. 
Accordingly, we rescale the whole hot spot such that its numerical integral does equal 1.
This ensures that, after generating the $\Nsrc$ sources, the numerical integral of $\Phi(\vb{x})$ over the whole grid is equal to $\Nsrc$, in accordance with Eq.~\eqref{eq:int_transverse_dens_IHSM}.

The final step is to recenter the generated initial-state profile, by translating it across the grid such that the profile center coincides with that of the grid up to less than $a/2$ along the $x$- and $y$-directions.
As already stated above, this induces (long-range) correlations between the hot spots~\cite{Blaizot:2014wba}, which are however small for the setups we consider.
In fact, the overlap between neighboring finite-size hot spots already induces correlations. 

This last source of correlations disappears if the sources are pointlike and their positions are sampled independently of each other. 
Such a scenario is however difficult to simulate with a discretized space: with a continuous distribution like Eq.~\eqref{eq:IPSM_f_1}, the probability that a pointlike hot spot falls on a grid site is essentially zero!
To still mimic the model, which will be referred to as $\ssrc=0$, we modify the implementation described above: 
After having randomly chosen a hot-spot position $\vb{x}_k$ with $f_1(\vb{x})$, we relocate the hot spot to the nearest grid point.\footnote{The probability that the two nearest neighboring sites are at exactly the same distance is again vanishing.}
Then we assign the value $1/a^2$ as the local value of the source function, to ensure that its numerical integral equals 1.
It is important to note that in this adjustment process, we inevitably introduce a small source of correlations. 
Yet, we view it as an acceptable compromise to be able to investigate numerically the model with pointlike hot spots, in which it is easier to gain also intuitive knowledge.

\section{Results}
\label{s:results}

Let us now give further details of our numerical simulations of the IHSM with different parameter sets (Sec.~\ref{ss:simulation_setup}).
We then present results for the average initial state (Sec.~\ref{ss:results_av-state}), followed by the fluctuation modes and their respective importance (Sec.~\ref{ss:results_modes}).
Eventually, we attempt a comparison with results from a MC Glauber model, by performing IHSM simulations with a different underlying hot-spot distribution (Sec.~\ref{ss:results_IHSM_vs_MCGlauber}).

\subsection{Simulation setup}
\label{ss:simulation_setup}

In the present study of the IHSM, we performed several ``runs'' of numerical simulations of the model, using different sets of parameters listed at the end of Sec.~\ref{ss:model_th}, to investigate their respective influence on the results.
These parameter sets are comprehensively summarized in Table~\ref{tab:sim_parameters}.
Throughout the paper, we use the parameters to label the results shown in the figures in the form $\mathrm{IHSM}(\sigma_x,\sigma_y)^{\ssrc}_{\Nsrc.}$ (omitting the unit for the widths) for the runs with fixed values of the hot-spot size $\ssrc$ and number $\Nsrc$, and with a uniform weight $\varpi=1$ for all sources.
Additionally, when no ambiguity arises, we dispense with explicitly indicating the values of $(\sigma_x,\sigma_y)$.

\begin{table}[!t]
\caption{\label{tab:sim_parameters} Summary of simulation parameters.}
\begin{ruledtabular}
\begin{tabular}{lcccc}
Parameter set & $\Nsrc$ & $\ssrc$ [fm] & $\sigma_x$ [fm] & $\sigma_y$ [fm]\\
\hline
IHSM(4,4)$_{50}^{0.3}$ & 50 & 0.3 & 4 & 4 \\
IHSM(4,4)$_{250}^{0.3}$ & 250 & 0.3 & 4 & 4 \\
IHSM(4,4)$_{750}^{0.3}$ & 750 & 0.3 & 4 & 4 \\
IHSM(4,4)$_{50}^{0.7}$ & 50 & 0.7 & 4 & 4 \\
IHSM(4,4)$_{250}^{0.7}$ & 250 & 0.7 & 4 & 4 \\
IHSM(4,4)$_{750}^{0.7}$ & 750 & 0.7 & 4 & 4 \\
IHSM(4,4)$_{250}^{0.0}$ & 250 & 0.0 & 4 & 4 \\
\hline
IHSM(2,4)$_{50}^{0.3}$ & 50 & 0.3 & 2 & 4 \\
IHSM(2,4)$_{250}^{0.3}$ & 250 & 0.3 & 2 & 4 \\
IHSM(2,4)$_{50}^{0.7}$ & 50 & 0.7 & 2 & 4 \\
IHSM(2,4)$_{250}^{0.7}$ & 250 & 0.7 & 2 & 4 \\
\hline
$\Nsrc = 50\pm 10$ & $50\pm10$ & 0.3 & 4 & 4 \\
$\varpi = 1\pm 0.3$ & 50 & 0.3 & 4 & 4 \\
$\sigma_{\rm src} = 0.3\pm 0.17$ & 50 & $0.3\pm 0.17$ & 4 & 4 \\
\end{tabular}
\end{ruledtabular}    
\end{table}

To fix ideas, the number of hot spots $\Nsrc$ can be compared to that of binary nucleon-nucleon ($N\hspace{-.1em}N$) collisions in a MC Glauber model. 
Using an inelastic $N\hspace{-.1em}N$ cross section of 67.6~mb for Pb-Pb collisions at $\sqrt{s_{_{N\hspace{-.1em}N}}}=5.02$~TeV, the values $\Nsrc=\{ 50, 250, 750\}$ correspond to collisions at impact parameters $b\approx \{ 12.5, 10, 7.5\}$~fm, respectively,\footnote{This can be for instance read off Fig.~(2.1) in Ref.~\cite{Roch:2022ofo}.}  i.e.\ to peripheral to semi-peripheral centralities.

Besides the runs with fixed parameters, we also performed three runs with a rotationally symmetric $f_1$ --- i.e.\ $\sigma_x=\sigma_y=4$~fm --- in which we let either the number of sources per initial state, or their weight $\varpi$, or the widths of the hot spots fluctuate, displayed in the last three lines of Table~\ref{tab:sim_parameters}. 
In all three cases, the fluctuating quantity was drawn from a uniform distribution over the interval specified in the Table. 

Each run consists of $\Nev = 2^{21}$ initial states.
In each run, we computed the average state $\bar{\Psi}(\vb{x})$, according to Eq.~\eqref{eq:avg_state}, and the fluctuation modes $\{\Psi_l(\vb{x})\}$ with their relative weights $\{w_l\}$, to which we shall come back in Sec.~\ref{ss:results_modes}.

\subsection{Average initial state}
\label{ss:results_av-state}

\begin{figure}[!t]
\includegraphics[width=\linewidth]{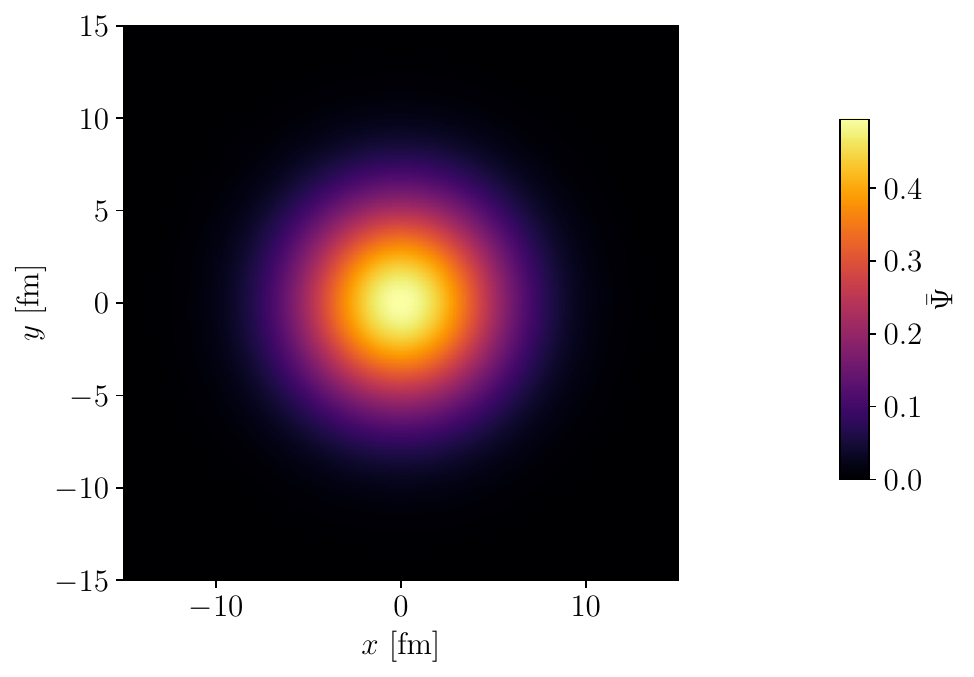}
\caption{Density profile of the average initial state $\bar{\Psi}(\vb{x})$ for the run IHSM(4,4)$_{50}^{0.3}$.}\vspace{-2mm}
\label{fig:av-state}
\end{figure}

As an example, we display in Fig.~\ref{fig:av-state} the average-state profile for the simulation consisting of configurations with $\Nsrc = 50$ hot spots with width $\ssrc=0.3$~fm, whose centers are distributed across a symmetric $f_1(\vb{x})$.
The profiles look very similar for all other runs with $\sigma_x=\sigma_y$, while for the runs with $\sigma_x\neq\sigma_y$ we obtain elongated shapes (not shown), as could be expected.

Equation~\eqref{eq:avg_state2a} describing the average initial state $\bar{\Psi}(\vb{x})$ is readily integrated over the transverse position $\vb{x}$: 
Since both the source function $s$ and the hot-spot-center distribution $f_1$ are normalized to 1, integrating $\bar{\Psi}(\vb{x})$ yields $\expval{\Nsrc}$. 
We checked that this is indeed the case of the average profile from our simulations in the various runs, up to numerical precision. 

When the source function has a fixed width $\ssrc$, since both $s(\vb{x})$ and the distribution of hot-spot positions $f_1(\vb{x})$ are Gaussian, a straightforward calculation with Eq.~\eqref{eq:avg_state2a} shows that $\bar{\Psi}$ is also a two-dimensional Gaussian, normalized to $\expval{\Nsrc}$, with widths $(\sigma_x^2+\ssrc^2)^{1/2}$, $(\sigma_y^2+\ssrc^2)^{1/2}$:
\begin{align}
\bar{\Psi}(\vb{x}) = &\frac{\expval{\Nsrc}}{2\pi\sqrt{(\sigma_x^2+\ssrc^2)(\sigma_y^2+\ssrc^2)}} \cr
&\times \exp\left[-\frac{x^2}{2(\sigma_x^2+\ssrc^2)}-\frac{y^2}{2(\sigma_y^2+\ssrc^2)}\right],
\label{eq:avg_state2} 
\end{align}
where we discard the correction due to the recentering of the profiles, which effectively leads to a small decrease of $\sigma_x$ and $\sigma_y$. 
That is, the average initial state actually extends over a (slightly) larger region than $f_1(\vb{x})$, due to the finite size of the hot spots themselves. 
This is most obvious in the extreme case of exactly localized hot-spot centers, $f_1(\vb{x}) = \delta^{(2)}(\vb{x})$, in which case the average state becomes proportional to the source function $s(\vb{x})$ and thus has the same width.

To check how the numerical simulations approach this formula, we computed the mean square widths $\{x^2\}_{\bar{\Psi}}$, $\{y^2\}_{\bar{\Psi}}$ of the average initial states of the various runs, where the curly brackets $\{\cdots\}_{\bar{\Psi}}$ denote an average weighted with $\bar{\Psi}$.
In the analytical case, Eq.~\eqref{eq:avg_state2}, these quantities respectively equal $\sigma_x^2+\ssrc^2$ and $\sigma_y^2+\ssrc^2$.
Numerically, we find a consistent trend over all runs:
At a given geometry of the hot-spot distribution, i.e.\ ultimately a given $\sigma_x$, the mean square widths $\{x^2\}_{\bar{\Psi}}$ and $\{y^2\}_{\bar{\Psi}}$ only depend on $\ssrc$, increasing with the source size, but they are independent (to two-digit accuracy, see below) of $\Nsrc$ and the possible inclusion of source weights.
The absence of a sizable dependence on the number of hot spots shows that the correction to Eq.~\eqref{eq:avg_state2} due to recentering is indeed invisible within the numerical accuracy of our simulations.

More precisely, for hot spots with $\ssrc=0.3$~fm resp.\ 0.7~fm, we obtain $\{y^2\}_{\bar{\Psi}}^{1/2} = 4.00$~fm resp.\ 4.03~fm \vspace{-.8mm} --- and the same $\{x^2\}_{\bar{\Psi}}^{1/2}$ values when $\sigma_x=\sigma_y$.
These results should be compared with the respective ``theoretical'' values $(\sigma_y^2+\ssrc^2)^{1/2} = 4.01$~fm or 4.06~fm. \vspace{-.7mm}
In the runs with $\sigma_x = 2$~fm, the simulations yield $\{x^2\}_{\bar{\Psi}}^{1/2} = 2.03$~fm resp.\ 2.12~fm for $\ssrc=0.3$~fm resp.\ 0.7~fm, to be compared with anticipated values $(\sigma_x^2+\ssrc^2)^{1/2} = 2.02$~fm resp.\ 2.12~fm.
That is, the numerical values are almost systematically smaller than those from Eq.~\eqref{eq:avg_state2}, but one can convince oneself that this arises from the discrete grid: 
The numerical version of the integral uses the values at the grid points, instead of the continuous range of values between successive grid points. 
Accordingly, it puts more weight on the points closest to the center, where $\bar{\Psi}(\vb{x})$ is larger, resulting in smaller mean-square widths.
That being said, it is remarkable that the discrepancy between the numerical and analytical values is at least a factor 8 smaller than the grid spacing.

In the case of pointlike sources, for which we performed a single run with a slightly different recipe, we obtain $\{x^2\}_{\bar{\Psi}}^{1/2} = \{y^2\}_{\bar{\Psi}}^{1/2} =  3.97$~fm, smaller than the expected value of 4~fm. 
The discrepancy is again due to the space discretization but of a different origin. 
In this scenario, the hot-spot contributions are shifted to the closest grid point, which has a higher probability of being closer to the center because $f_1(\vb{x})$ decreases with distance from the center. 
Thus, there is a higher likelihood of a pointlike source being moved closer to the center, again resulting in smaller mean square widths.

\subsection{Mode-by-mode decomposition of the IHSM}
\label{ss:results_modes}

To determine the fluctuation modes $\{\Psi_l(\vb{x})\}$ such that the expansion coefficients $c_l$ entering the decomposition~\eqref{eq:evt-fluct_vs_modes} of the fluctuations $\delta\Phi^{(i)}(\vb{x})$ are centered and uncorrelated, according to Eq.~\eqref{eq:mean_and_variance_cl2}, the recipe introduced in Ref.~\cite{Borghini:2022iym} is to define numerically a ``density matrix'' $\rho$ as
\begin{equation}
\rho \equiv \frac{1}{\Nev}\sum_i  \Phi^{(i)} \Phi^{(i)\mathsf{T}} - \bar{\Psi}\bar{\Psi}^\mathsf{T},
\label{eq:rho}
\end{equation}
where the initial states $\{\Phi^{(i)}(\vb{x})\}$ are represented by their components on some arbitrary finite basis --- which in practice we take to be the trivial basis associated with the grid.
Since there are $N_s^2$ such basis vectors, $\rho$ is a $(N_s^2\times N_s^2)$-dimensional matrix.
The searched-for fluctuation modes $\{\Psi_l\}$ are then eigenvectors of $\rho$ and their squared norms $\{\lambda_l\}$ are the respective eigenvalues.

\subsubsection{Modes}

\begin{figure*}
\includegraphics[height=0.975\textheight]{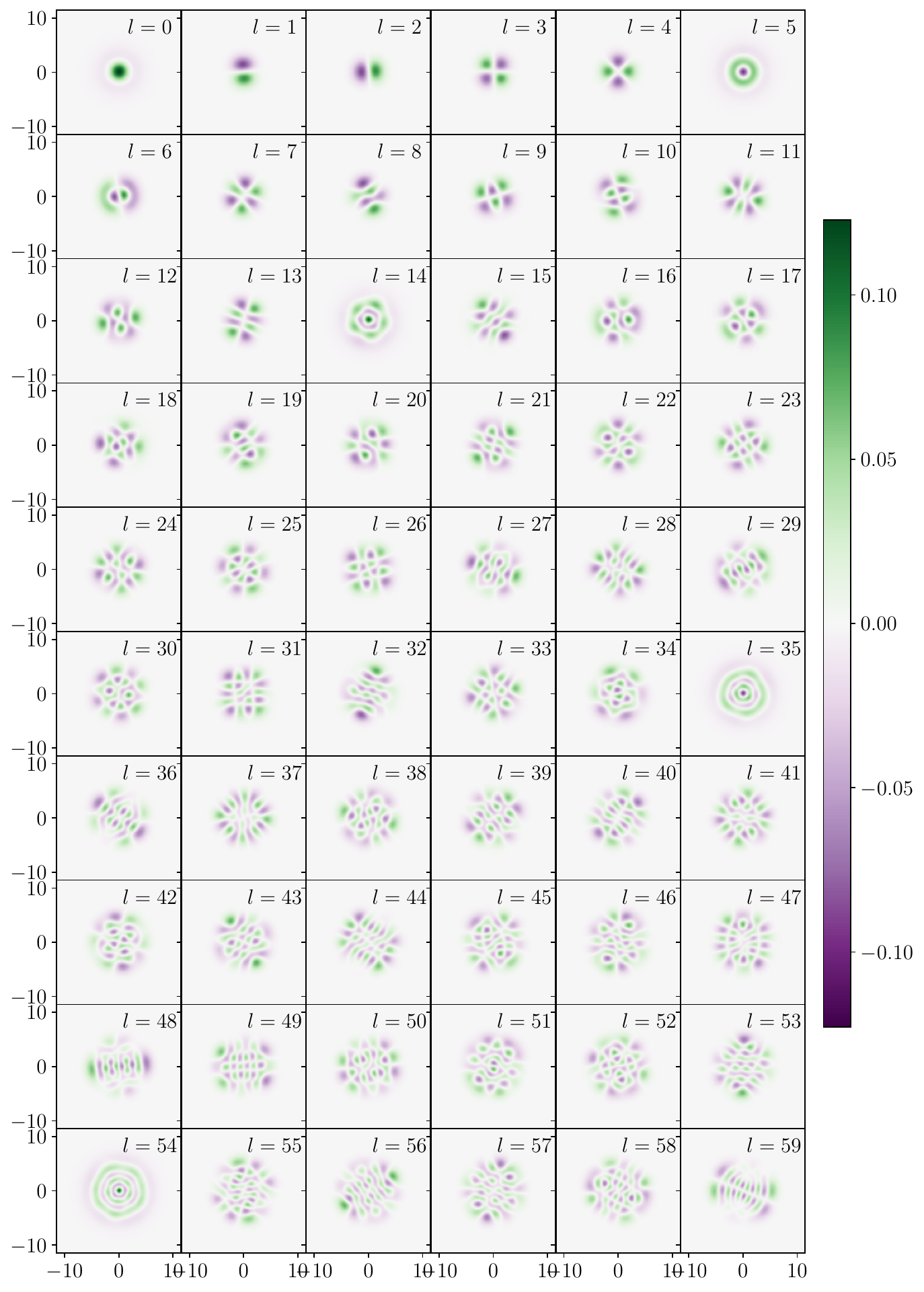}
\vspace{-3mm}
\caption{Density plots of the first 60 orthonormal eigenvectors for the run IHSM(4,4)$_{50}^{0.3}$. Both axes are in units of fm.}
\label{fig:Modes-example}
\end{figure*}

The modes themselves, and how their properties affect final-state observables at the end of a dynamical evolution, are not within the scope of the present study, since the model we used to generate the initial profiles is at most semi-realistic. 
Nevertheless, we show as an example in Fig.~\ref{fig:Modes-example} the first sixty corresponding normalized eigenvectors $\{\Psi_l/\sqrt{\lambda_l}\}$ for the run IHSM(4,4)$_{50}^{0.3}$, where the modes are ordered by decreasing $\lambda_l$. 
The eigenvectors look similar to those shown in Ref.~\cite{Borghini:2022iym} for realistic initial-state models: there are modes with rotational symmetry, like the average initial state in Fig.~\ref{fig:av-state}; modes with more or less clearly recognizable dipole, quadrupole, hexapole\dots\ shape, which come in pairs; and modes with more complicated shapes. 

In comparison to the modes reported in Ref.~\cite{Borghini:2022iym}, there are two noteworthy differences. 
First, here we present fluctuation modes for initial states with only 50 hot spots, while the central MC Glauber events of Ref.~\cite{Borghini:2022iym} typically correspond to about $N_\textrm{coll.}\approx 2000$ binary collisions, and thus at least as many ``hot spots''.
The second difference is that in the initial states considered in run IHSM(4,4)$_{50}^{0.3}$ (and in almost all runs of Table~\ref{tab:sim_parameters}), the number of hot spots is fixed.
Assuming momentarily to facilitate the discussion that each source deposits energy, a fixed $\Nsrc$ means (as long as all hot spots contribute the same) a fixed total energy in the initial profile. 
We mentioned above that the average initial state $\bar{\Psi}$ is normalized to $\Nsrc$, i.e.\ it contains precisely the same energy. 
Now since the fluctuation modes are uncorrelated with each other, they cannot contribute energy, otherwise, any energy excess or deficit due to the presence of one mode would have to be exactly compensated by other modes, which would induce correlations. 
Mathematically, this requirement translates into the integral of every $\Psi_l(\vb{x})$ over position being zero.
We checked numerically that this is the case --- to be more precise, the absolute value of the integral is always smaller than $10^{-3}$, to be compared with the value $\Nsrc$ of the integral of $\bar{\Psi}$ --- for the modes in the runs with a fixed number of hot spots and uniform weights $\varpi$.
In the two runs with either fluctuating $\Nsrc$ or fluctuating $\varpi$, a few modes do have a non-zero integral, i.e.\ contribute some energy, namely those with rotational symmetry.

Eventually, another important ingredient of the mode-by-mode decomposition is the conditions~\eqref{eq:mean_and_variance_cl}, or more abstractly Eq.~\eqref{eq:mean_and_variance_cl2}, on the expansion coefficients $\{c_l\}$ of individual fluctuations over the basis of fluctuation modes.
For the sake of completeness, we present a few results on the statistics of these coefficients in Appendix~\ref{s:c_l}.

\subsubsection{Eigenvalues}

Let us now discuss the eigenvalues $\{\lambda_l\}$ of the density matrix $\rho$, or equivalently the relative weights $\{w_l\}$ of the modes, defined in Eq.~\eqref{eq:w_l}.
In essence, the density matrix is a discretized version of 
\begin{equation}
\rho(\vb{x},\vb{y}) \equiv 
\frac{1}{\Nev}\sum_i \Phi^{(i)}(\vb{x})_{}\Phi^{(i)}(\vb{y}) - \bar{\Psi}(\vb{x})\bar{\Psi}(\vb{y}).
\end{equation}
Since $\bar{\Psi}(\vb{x})$ is nothing but the average of the $\{\Phi^{(i)}(\vb{x})\}$, see Eq.~\eqref{eq:avg_state},  $\rho(\vb{x},\vb{y})$ is actually the (auto)correlation function of the fluctuations $\{\delta\Phi^{(i)}(\vb{x})\}$. 
Viewing the latter as realizations of a random function $\delta\Phi(\vb{x})$, one has
\begin{equation}
\label{eq:rho(x,y)}
\rho(\vb{x},\vb{y}) = \expval{\delta\Phi(\vb{x})_{}\delta\Phi(\vb{y})}.
\end{equation}
This is the function denoted by $S(\vb{x},\vb{y})$ in Ref.~\cite{Blaizot:2014wba}.
Invoking Eq.~(C3) of that article\footnote{The original equation has a misprint: the two-point density $f_2(\vb{x}_1,\vb{x}_2)$ should be multiplied by a factor $N(N-1)$ instead of $N$ only.}, it becomes
\begin{widetext}
\begin{align}
\rho(\vb{x},\vb{y}) =
&\expval{\Nsrc}\!\expval{\varpi^2}\!\!\int\!s(\vb{x}-\vb{x}_1)_{}s(\vb{y}-\vb{x}_1)_{}f_1(\vb{x}_1)\dd[2]{\vb{x}_1} \cr 
&+ \expval{\Nsrc(\Nsrc-1)}\!\expval{\varpi}^{\!2}\!\int\!s(\vb{x}-\vb{x}_1)_{}s(\vb{y}-\vb{x}_2)_{}f_2(\vb{x}_1,\vb{x}_2)\dd[2]{\vb{x}_1}\dd[2]{\vb{x}_2} - \bar{\Psi}(\vb{x})\bar{\Psi}(\vb{y}),
\label{eq:rho(x,y)_IHSM}
\end{align}
\end{widetext}
where $f_2$ denotes the two-point distribution function of hot spots.
For independent sources, the latter factorizes into the product of the single-point distributions: $f_2(\vb{x}_1,\vb{x}_2) = f_1(\vb{x}_1)f_1(\vb{x}_2)$, which we shall assume from now on.  
We know that recentering induces correlations, yet from our findings for the average state $\bar{\Psi}$ we are confident that they are very small in our numerical simulations. 
With this factorization assumption, the integral involving $f_2$ in Eq.~\eqref{eq:rho(x,y)_IHSM} equals $\bar{\Psi}(\vb{x})\bar{\Psi}(\vb{y})/\!\expval{\Nsrc}^{\!2}$ thanks to Eq.~\eqref{eq:avg_state2a}.
Using $\expval{\varpi} = 1$, one eventually finds
\begin{align}
\rho(\vb{x},\vb{y}) =
&\expval{\Nsrc}\!\expval{\varpi^2}\!\!\int\!\!s(\vb{x}-\vb{x}_1)_{}s(\vb{y}-\vb{x}_1)_{}f_1(\vb{x}_1)\dd[2]{\vb{x}_1}\quad\cr
& -\frac{\expval{\Nsrc}-\sigma_{\Nsrc}^2}{\expval{\Nsrc}^{\!2}} \bar{\Psi}(\vb{x})\bar{\Psi}(\vb{y}),
\label{eq:rho(x,y)_independent}
\end{align}
where we introduced the variance $\sigma_{\Nsrc}^2$ of the fluctuations in the number of hot spots.
With Gaussian $s(\vb{x})$ --- i.e.\ for finite-size hot spots --- and $f_1(\vb{x})$, the first term on the right-hand side is also Gaussian. 
We do not show the exact form of $\rho(\vb{x},\vb{y})$ in the general case, since it is not used in the following. 
Yet, it is worth noting that within the IHSM, it is entirely determined by the few parameters of the model. 

The second line of Eq.~\eqref{eq:rho(x,y)_independent} is somewhat misleading: 
since $\bar{\Psi}(\vb{x})$ is proportional to $\expval{\Nsrc}$, see Eq.~\eqref{eq:avg_state2a}, the actual dependence of the term on the number of hot spots is given by the numerator, i.e.\ it scales like $\expval{\Nsrc}-\sigma_{\Nsrc}^2$. 
When the number of sources is constant, so that $\sigma_{\Nsrc}=0$, the two-point correlation function $\rho(\vb{x},\vb{y})$ is thus proportional to $\Nsrc$. 
This should also hold, up to numerical fluctuations, for its discretized version~\eqref{eq:rho}. 
That is, one can expect that the eigenvalues $\{\lambda_l\}$ at fixed source-size $\ssrc$ and geometry of the hot-spot distribution should be proportional to $\Nsrc$. 

\begin{figure}[!t]
\includegraphics[width=\linewidth]{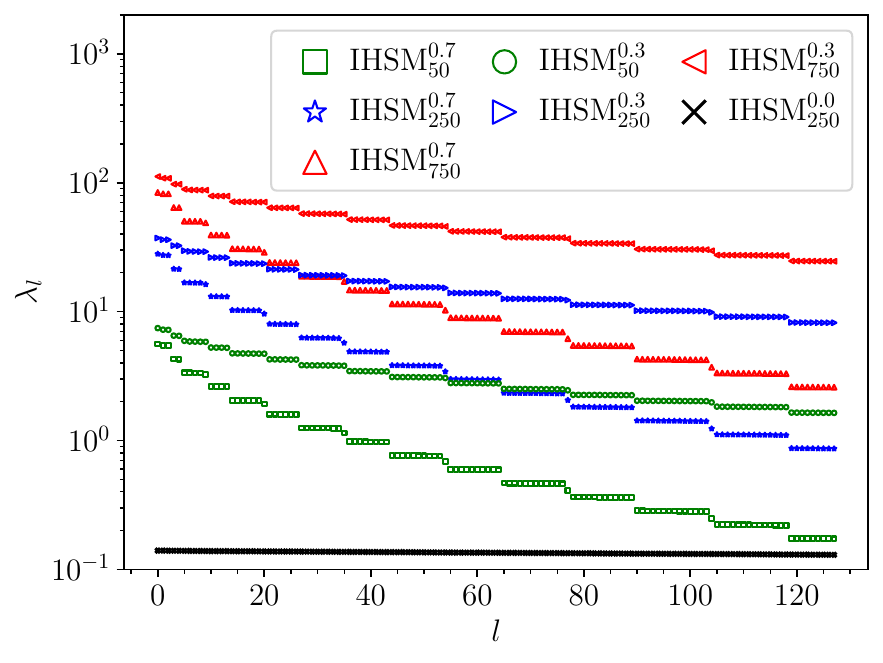}
\includegraphics[width=\linewidth]{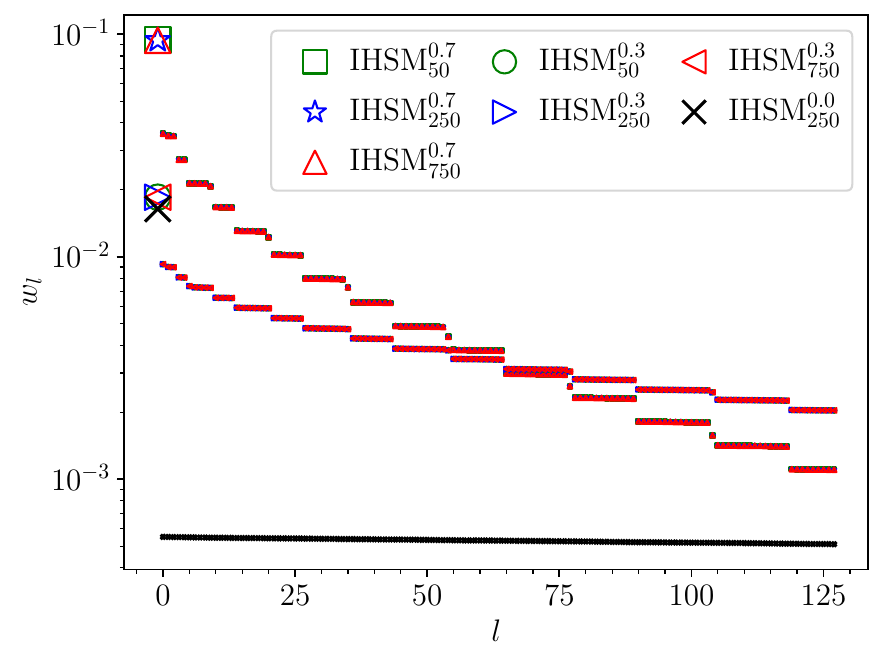}\vspace{-2mm}
\caption{Eigenvalues $\lambda_l$ (top) and relative weights $w_l$ (bottom) of the fluctuation modes in simulations with a rotationally symmetric hot-spot distribution. The large marker at $l=-1$ in the lower plot indicates the weight $\bar{w}$ of the average state.}
\label{fig:eigenvalues1}
\end{figure}

In Fig.~\ref{fig:eigenvalues1} we show the eigenvalues (top) and the relative weights $\{w_l\}$ of the fluctuation modes (bottom) for the seven first runs of Table~\ref{tab:sim_parameters}, i.e.\ the simulations with a rotationally symmetric hot-spot distribution and no fluctuation in the other parameters $\Nsrc$, $\varpi$ or $\ssrc$.

Let us leave momentarily aside the results of the run with pointlike sources. 
For each finite value of $\ssrc$, the three runs with different $\Nsrc$ yield $\{\lambda_l\}$ spectra that look extremely similar, up to a multiplicative factor, which according to our previous reasoning should be $\Nsrc$.
When going to the relative weights $\{w_l\}$, which are proportional to the square roots of the eigenvalues, we expect to cancel out the dependence on $\Nsrc$, since both numerator and denominator of Eq.~\eqref{eq:w_l} scale like $\sqrt{\Nsrc}$. 
This is indeed what can be observed in the lower panel, in which the $\{w_l\}$ spectra for runs with the same hot spot size $\ssrc$ and different $\Nsrc$ collapse together. 

Discarding the trivial dependence on the number of sources, we can now focus on other features of the eigenvalues across different hot-spot sizes $\ssrc$. 
The spectra display distinct steps, corresponding to (almost) degenerate fluctuation modes. 
This was already observed in Ref.~\cite{Borghini:2022iym}: as discussed there, it is, for instance, clear that the two ``dipole'' modes $l=1$ and 2 of Fig.~\ref{fig:Modes-example}, that differ only by a rotation by $\pi/2$, should be exactly degenerate in a model with rotational symmetry. 
A novelty here is when we compare simulations at different $\ssrc$: the quasi-degeneracy steps seem to be similar --- with the same length, possibly up to one mode --- at the two values of the hot-spot size.
We have no explanation for this feature, which is all the most remarkable when we come to the last point, namely the significant difference in the drop-off of the spectra according to the hot-spot size. 

Indeed, one sees that the spectra of eigenvalues or relative weights are steeper for simulations with larger hot spots ($\ssrc=0.7$~fm) than for the runs with smaller ones ($\ssrc=0.3$~fm), while the spectra become almost flat for pointlike sources.%
\footnote{To correct the visual impression, let us quote a few values: with pointlike hot spots, the relative weight is $w_0 \simeq 0.551\times 10^{-3}$ for the ``dominant'' mode, $w_{127} \simeq 0.511\times 10^{-3}$ for the last mode shown in Fig.~\ref{fig:eigenvalues1}, and $w_{255} \simeq 0.476\times 10^{-3}$ for the 256th mode.}
In parallel, one sees in the lower panel of Fig.~\ref{fig:eigenvalues1} that the relative weight $\bar{w}$ of the average initial state is higher (approximately 10\%) for the runs with larger sources, than for those with smaller hot spots (about 2\% for $\ssrc=0.3$~fm and slightly less for pointlike sources).
This means that higher fluctuation modes $\Psi_l$ become less and less important when the hot-spot size grows. 
Now, in the IHSM (with finite $\ssrc$) of the present study as well as the two more realistic models of Ref.~\cite{Borghini:2022iym}, the modes with higher index $l$ are those with structure on increasingly shorter length scales. 
That such fluctuation modes contribute less when the source size increases, and thus induce correlations of increasing wavelength, is not surprising. 
In turn, the almost flat spectrum in the simulations with pointlike hot spots suggests that fluctuations at all (subnuclear) length scales are almost equally important in that scenario. 

\begin{figure}[!t]
\includegraphics[width=\linewidth]{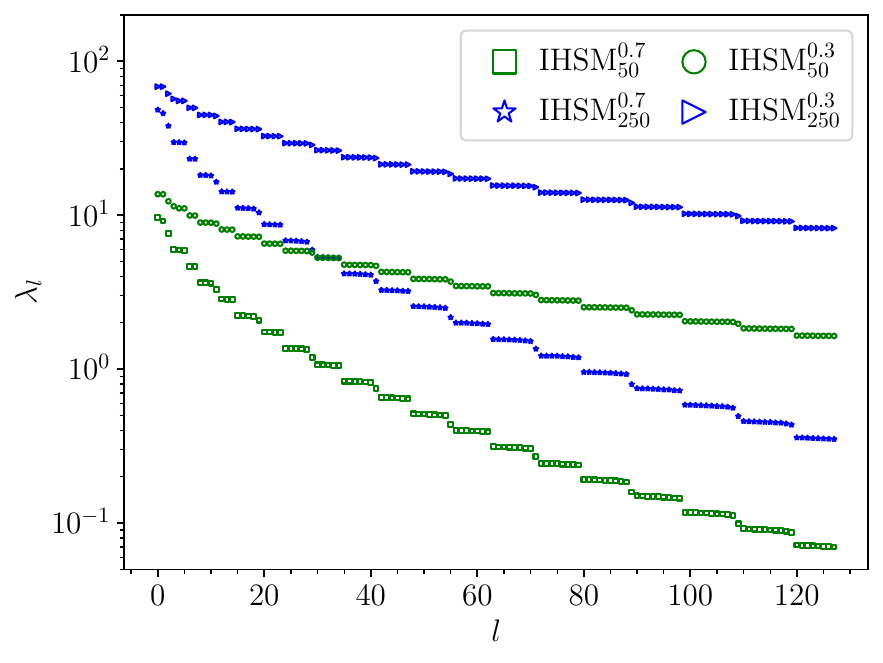}
\includegraphics[width=\linewidth]{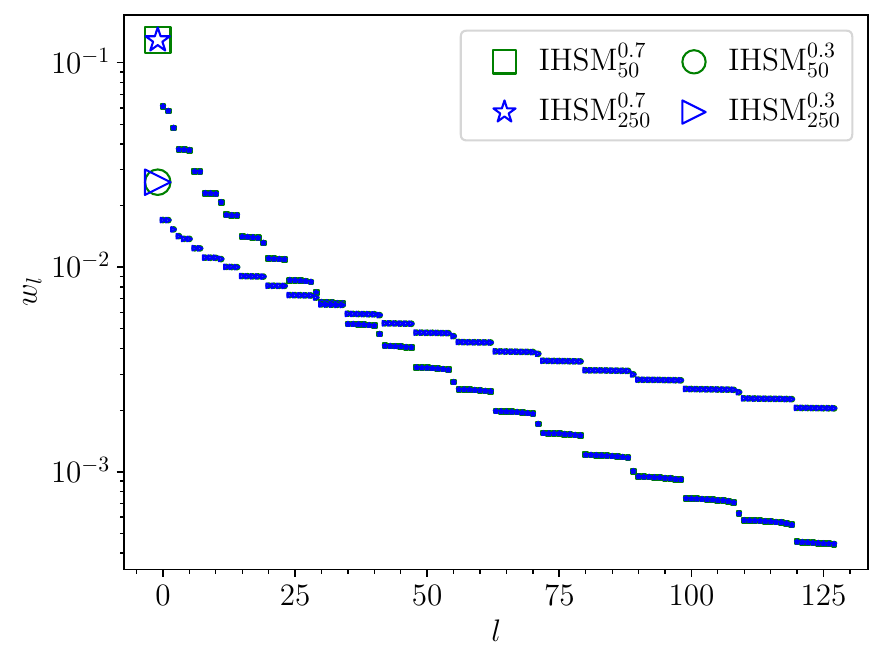}\vspace{-2mm}
\caption{Same as Fig.~\ref{fig:eigenvalues2} for the simulations with an elliptic hot-spot distribution.}
\label{fig:eigenvalues2}
\end{figure}

Similar trends are found in the spectra of eigenvalues and relative weights, shown in Fig.~\ref{fig:eigenvalues2}, in simulations with elongated hot-spot distribution.
Here as well, the dependence on the number of sources is an overall factor in the $\{\lambda_l\}$, that disappears when going to the $\{w_l\}$. 
And again one finds that the spectra are steeper for larger hot spots. 
A small difference with the results of Fig.~\ref{fig:eigenvalues1} is that the ``quasi-degenerate'' plateaus generally involve fewer fluctuation modes. 
This is consistent with the fact that rotational symmetry is now explicitly broken so that the degeneracy between pairs of modes with the same geometry rotated by a fraction of $\pi$ is lifted.

Before we present the results of the simulations in which we let one of the parameters fluctuate, let us come back to the two-point correlation function.
Besides the scaling behavior with the number of hot spots, one can further exploit Eq.~\eqref{eq:rho(x,y)_independent} in its generality for a special case, namely $\vb{x}=\vb{y}$. 
Indeed, the equation gives at once $\rho(\vb{x},\vb{x})$, which can then be integrated over all transverse positions $\vb{x}$.
With finite-size Gaussian hot spots, the calculation is straightforward: 
The squared profile function~\eqref{eq:s(x)} is Gaussian with width $\ssrc/\sqrt{2}$ and normalized to $1/4\pi\ssrc^2$. 
Integrating first over $\vb{x}$ gives this normalization, and the integral over $\vb{x}_1$ in Eq.~\eqref{eq:rho(x,y)_independent} is then trivial. 
In turn, $\bar{\Psi}(\vb{x})^2$ is also Gaussian, normalized to $\expval{\Nsrc}^{\!2\!}/[4\pi(\sigma_x^2+\ssrc^2)^{1/2}(\sigma_y^2+\ssrc^2)^{1/2}]$.
All in all, one thus finds
\begin{align}
\int\!&\rho(\vb{x},\vb{x})\dd[2]{\vb{x}} = \frac{\expval{\Nsrc}}{4\pi\ssrc^2} \cr
&\times\!\Bigg[ \!\expval{\varpi^2} - \bigg(\!1-\frac{\sigma_{\Nsrc}^2}{\expval{\Nsrc}}\!\bigg)
\frac{\ssrc^2}{\sqrt{(\sigma_x^2\!+\!\ssrc^2)(\sigma_y^2\!+\!\ssrc^2)}}\Bigg].\quad\ \ 
\label{eq:int_rho(x,x)}
\end{align}
Note that this expression diverges in the limit $\ssrc\to 0$ of pointlike sources. 
This can easily be traced back to the fact that the term $s(\vb{x}-\vb{x}_1)s(\vb{y}-\vb{x}_1)$ in the integrand of Eq.~\eqref{eq:rho(x,y)_independent} becomes an ill-defined squared (two-dimensional) $\delta$-distribution when $\vb{x}=\vb{y}$. 

The interest of the integral~\eqref{eq:int_rho(x,x)} is that it represents, up to a factor $a^2$ corresponding to the area of an elementary cell of our spatial grid, the trace of the matrix $\rho$:
\begin{equation}
\int\!\rho(\vb{x},\vb{x})\dd[2]{\vb{x}} = \Tr(\rho)\,a^2.
\label{eq:int_rho(x,x)_vs_Tr}
\end{equation}
Now, in our numerical simulations, we can clearly compute the trace of the matrix defined by Eq.~\eqref{eq:rho}, yielding a first determination of this trace, which we denote by $\Tr(\rho)_\textrm{num.}$. 
We checked that summing over the $N_s^2$ computed eigenvalues gives consistent values. 
On the other side, we can also calculate the analytical prediction for the trace using Eq.~\eqref{eq:int_rho(x,x)} with the input parameters of the simulations and Eq.~\eqref{eq:int_rho(x,x)_vs_Tr} with the known grid spacing: 
this yields a value $\Tr(\rho)_\textrm{an.}$. 
We list the corresponding values for our different runs in Table~\ref{tab:trace_density_matrix}, including the scenarios that we shall discuss shortly. 
\begin{table}[!t]
\caption{\label{tab:trace_density_matrix} Trace of the density matrix $\rho$ computed from the analytical expression Eqs.~\eqref{eq:int_rho(x,x)}--\eqref{eq:int_rho(x,x)_vs_Tr} and from the numerical simulations.}
\begin{ruledtabular}
\begin{tabular}{lcc}
Parameter set & $\Tr(\rho)_\textrm{an.}$ & $\Tr(\rho)_\textrm{num.}$ \\ \hline
IHSM(4,4)$_{50}^{0.3}$ & 787.86 & 789.04 \\
IHSM(4,4)$_{250}^{0.3}$ & 3939.28 & 3945.51 \\
IHSM(4,4)$_{750}^{0.3}$ & 11817.83 & 11836.24 \\

IHSM(4,4)$_{50}^{0.7}$ & 141.20 & 141.96 \\
IHSM(4,4)$_{250}^{0.7}$ & 705.99 & 709.89 \\
IHSM(4,4)$_{750}^{0.7}$ & 2117.97 & 2129.28 \\

IHSM(4,4)$_{250}^{0.0}$ & divergent & 249.93 \\
\hline
IHSM(2,4)$_{50}^{0.3}$ & 783.50 & 784.71 \\
IHSM(2,4)$_{250}^{0.3}$ & 3917.49 & 3923.63 \\
IHSM(2,4)$_{50}^{0.7}$ & 137.24 & 137.97 \\
IHSM(2,4)$_{250}^{0.7}$ & 686.18 & 689.99 \\
\hline
$\Nsrc = 50\pm 10$ & 791.11 & 792.23 \\
$\varpi = 1 \pm 0.3$ & 811.62 & 812.94 \\
$\ssrc = 0.3 \pm 0.17$ & 1162.61 & 1184.57 \\
\end{tabular}
\end{ruledtabular}
\end{table}

The agreement between the analytically and numerically determined traces is generally excellent across all simulations, mostly at the 1\% level or better.
This makes us confident that the computer implementation of the toy model is not plagued by large numerical errors.
Interestingly, $\Tr(\rho)_\textrm{num.}$ is systematically larger than $\Tr(\rho)_\textrm{an.}$ (when defined).
We did not attempt to track the precise origin of the discrepancy between the two sets of values.
However, it is clear that the matrix $\rho$ should ideally contain a number of zero values on the diagonal --- for instance, at the points at the edges of our grid, at a distance of at least 15~fm from the center --- which are finite and positive in the numerical implementation, thereby biasing the trace.

As stated above, Eq.~\eqref{eq:int_rho(x,x)} yields a divergent result in the limit  $\ssrc\to 0$ of pointlike sources. 
This divergence cannot occur in the numerical determination, with a finite-dimensional density matrix $\rho$, whose diagonalization yields $N_s^2$ finite eigenvalues (and their eigenvectors) and thus a well-defined trace.

\begin{figure}[!t]
\includegraphics[width=\linewidth]{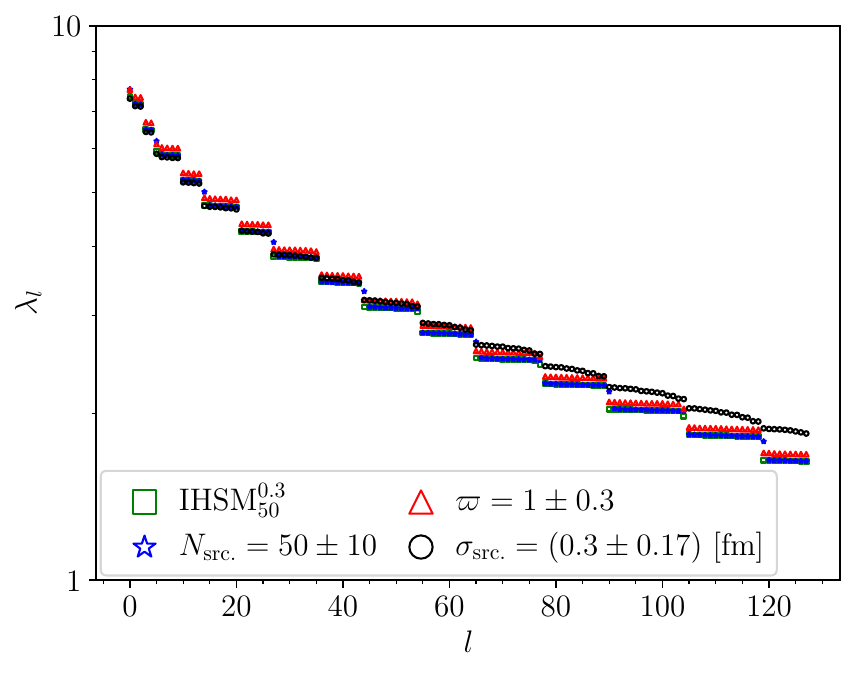}\vspace{-2mm}
\caption{Eigenvalues $\lambda_l$ for the simulations with rotationally symmetric hot-spot distribution, $\Nsrc=50$, $\ssrc=0.3$~fm, and unit source weights $\varpi=1$ (IHSM$^{0.3}_{50}$), and runs in which either $\Nsrc$, $\ssrc$, or $\varpi$ fluctuate about these values.}
\label{fig:eigenvalues4}
\end{figure}

Let us now discuss the three runs in which we let one of the model parameters fluctuate, either event-to-event (fluctuations in $\Nsrc$), or from one hot spot to the next (fluctuating weights $\varpi$ and source sizes $\ssrc$).
In Fig.~\ref{fig:eigenvalues4} we display the eigenvalues for the run with exactly $\Nsrc=50$ hot spots, unit weights, and $\ssrc=0.3$~fm (with a rotationally symmetric hot-spot distribution), and for the simulations in which we allow for uniformly distributed fluctuations about the values of this ``reference'' run. 

The run with fluctuations $\Nsrc = 50\pm 10$ in the hot-spot number\footnote{This results in $\sigma_{\Nsrc}^2=\Delta\Nsrc(1+\Delta\Nsrc)/3$ with $\Delta\Nsrc=10$ for Eqs.~\eqref{eq:rho(x,y)_independent} and \eqref{eq:int_rho(x,x)}.} yields fluctuation modes whose associated eigenvalues are almost the same as in the run with fixed $\Nsrc$, up to an important exception, namely the class of ``rotationally symmetric'' modes, which appear to have a consistently higher $\lambda_l$ when $\Nsrc$ is allowed to fluctuate. 
We have already mentioned that these specific modes have a non-vanishing integral over the transverse plane when $\Nsrc$ can fluctuate, in contrast to the fixed-$\Nsrc$ case. 
It is interesting to observe that this difference is accompanied by a sizable change in the relative importance of the modes.

Turning now to the initial states with hot spots with fluctuating weight, $\varpi = 1 \pm 0.3$,  the eigenvalue spectrum in Fig.~\ref{fig:eigenvalues4} appears to parallel exactly that of the run with a fixed weight, up to a constant multiplicative factor. 
Equation~\eqref{eq:int_rho(x,x)} shows that with $\sigma_x,\sigma_y>10\,\ssrc$, as in our simulations with symmetric $f_1(\vb{x})$, the term in $\langle \varpi^2\rangle$ in Eq.~\eqref{eq:rho(x,y)_independent} is typically larger than the other by a factor of order 100. 
This then results in eigenvalues of $\rho$ that approximately scale linearly with $\langle \varpi^2\rangle$, leading to the observed behavior in the present run where $\langle \varpi^2\rangle=1.03$, instead of 1 in the case of fixed unit weights. 

The profiles with fluctuations $\ssrc = (0.3 \pm 0.17)$~fm in the source size are somewhat special because the corresponding average initial state $\bar{\Psi}(\vb{x})$ slightly differs from the other runs, since the source function $s(\vb{x})$ entering Eq.~\eqref{eq:avg_state2} does not depend linearly on $\ssrc$. 
As $\ssrc$ remains significantly smaller than the width $\sigma_x=\sigma_y$ of the hot spot distribution, this change is however minimal, so we did not attempt to optimize the interval over which $\ssrc$ fluctuates to try and keep the properties of $\bar{\Psi}(\vb{x})$ unchanged.%
\footnote{For this run, we computed $\Tr(\rho)_\textrm{an.}$ in Table~\ref{tab:trace_density_matrix} by averaging numerically the trace values given by the analytical formula at fixed $\ssrc$.}
In Fig.~\ref{fig:eigenvalues4} one sees that these fluctuations lead to the greatest difference with the reference run, which is possibly not a surprise since Figs.~\ref{fig:eigenvalues1} and \ref{fig:eigenvalues2} already showed the marked influence of the hot-spot size. 
More precisely, we see that the ``quasi-degenerate steps'' of the eigenvalue spectrum are less flat than at fixed $\ssrc$, but are now sizably slanted, which means less degeneracy. 
In addition, the eigenvalues of the higher fluctuation modes are larger than in the reference run, i.e.\ those modes are comparatively more important. 
This can safely be attributed to the at times smaller hot spots generated in the present run, which result as above in more frequent, and thus more important, fluctuations with smaller wavelengths.

\subsection{Comparison of the IHSM with the MC Glauber model}
\label{ss:results_IHSM_vs_MCGlauber}

\begin{figure}[!t]
	\includegraphics[width=\linewidth]{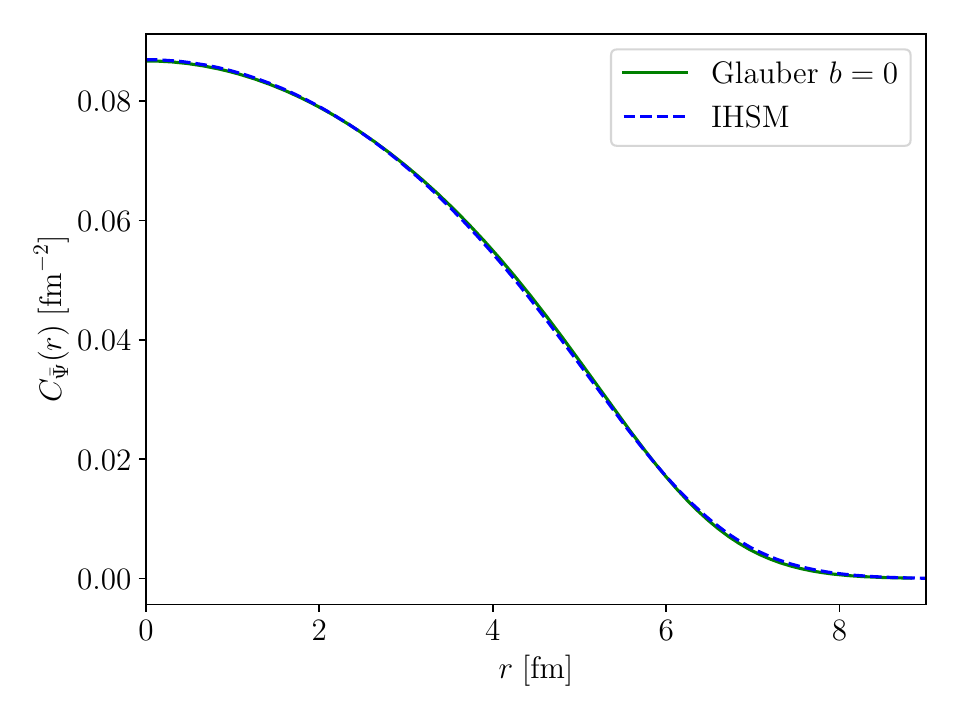}\vspace{-2mm}
	\caption{Transverse profile of the average initial state $\bar{\Psi}(\vb{x})$ (normalized to unity) for the IHSM (dashed) and the MC Glauber model (full line).}
	\label{fig:Profiles-Glauber}
\end{figure}

In Ref.~\cite{Blaizot:2014wba}, the eccentricity and size fluctuations computed for Pb-Pb collisions at $b=0$ in the MC Glauber from GLISSANDO~\cite{Broniowski:2007nz} were compared to results from the IHSM. 
For such observables, integrated over the whole transverse area of the initial state, a good agreement was found between both approaches, although the MC Glauber includes (intra)nuclear correlations. 
In this Section, we pursue the same idea and compare the fluctuation modes and respective relative weights $w_l$ from a MC Glauber code, already presented in Ref.~\cite{Borghini:2022iym}, with those from IHSM simulations with similar global characteristics. 
More specifically, we now used hot spots with a width $\ssrc = 0.4$~fm, which is the smearing radius used in the MC Glauber simulations~\cite{Borghini:2022iym}.
We also tuned the hot-spot-center distribution $f_1(\vb{x})$ in the IHSM such that the resulting average initial state $\bar{\Psi}(\vb{x})$ is the same as computed in the MC Glauber for Pb-Pb collisions at 5.02~TeV. 
This is shown in Fig.~\ref{fig:Profiles-Glauber}, in which the normalized radial profiles ${\cal C}_{\bar{\Psi}}(r)\equiv\bar{\Psi}(|\vb{x}|) / \norm{\bar{\Psi}}$ of $\bar{\Psi}(\vb{x})$ computed in both models are compared, and indeed almost coincide.

A significant difference between the parameters of the two sets of simulations we now compare is the number of hot spots.
In the MC Glauber model, it fluctuates event by event, with a mean value of order 2500 hot spots. 
In contrast, we simulated only $\Nsrc = 250$ sources in the IHSM, to accelerate the event generation. 
According to our findings in Sec.~\ref{ss:results_modes}, this difference in the (mean) number of sources should have a negligible impact on the relative weights $w_l$, since the eigenvalues $\lambda_l$ scale with $\Nsrc$. 
The only significant influence is that the fluctuations in $\Nsrc$ in the MC Glauber mean that the total energy in the initial states is not constant: as we argued above when looking at the IHSM with $\Nsrc = 50\pm 10$ sources, there should be a few modes that are responsible for these energy fluctuations, and the associated relative weight $w_l$ may differ from the case with fixed $\Nsrc$.

\begin{figure}[!t]
	\includegraphics[width=\linewidth]{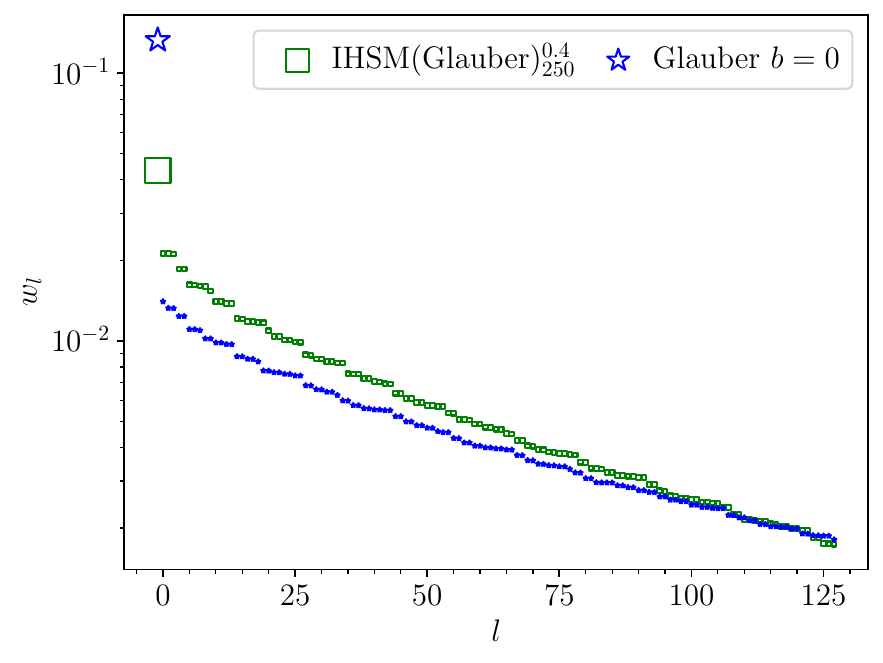}\vspace{-2mm}
	\caption{\label{fig:eigenvalues-Glauber}Relative weights $w_l$ of the fluctuation modes in simulations with the IHSM (stars) and the MC Glauber model (squares). The large markers at $l=-1$ indicate the weight $\bar{w}$ of the average state.}	
\end{figure}
The relative weights $w_l$ of the fluctuation modes for both simulations are shown in Fig.~\ref{fig:eigenvalues-Glauber}, where the results of the IHSM are labeled IHSM(Glauber)$_{250}^{0.4}$ to keep track of the parameters used to generate the initial states.\footnote{The fluctuation modes for the IHSM are shown in Fig.~\ref{fig:Modes-IHSM(Glauber)} in Appendix~\ref{s:modes_ISHM(Glauber)}, while those for the MC Glauber model can be found in Fig.~30 of Ref.~\cite{Borghini:2022iym}.} 
The spectrum for the IHSM is clearly steeper, although the difference in slope with the MC Glauber results is less marked than when changing the hot spot size from 0.3 to 0.7~fm in Fig.~\ref{fig:eigenvalues1} or Fig.~\ref{fig:eigenvalues2}.
The spectra really cross around mode $l\simeq 115$: beyond $l=125$, the relative weights for the MC Glauber model are larger than in the IHSM.
When looking in detail, one sees that the degeneracy patterns in both sets differ, although not much --- indeed, the steps that are clearly marked in the IHSM can also be recognized in the MC Glauber at the same $l$ values.
As in Fig.~\ref{fig:eigenvalues1}, the relative weight of the average state $\bar{\Psi}$ is about a factor 2 larger than the relative weight of the ``dominant mode'' $w_0$ in the IHSM, while they differ by a factor 10 in the MC Glauber initial states.

The latter observation is a clear hint that there is a significant difference between both models, even with similar parameters (apart from $\Nsrc$, whose role we believe to be unimportant).
There exist ``short range correlations'' in the MC Glauber model, that are not present in the IHSM.
An example is the repulsive core implemented for the positions of nucleons in the colliding nuclei. 
But it may possibly be more important that each nucleon-nucleon collision actually results in three correlated hot spots: one at each nucleon-center position at a maximal distance $\sqrt{\sigma_{_{\!N\hspace{-.1em}N}}/\pi}\simeq 1.47$~fm, and one inbetween.\footnote{The hot spots at the participant positions all contribute the same amount of energy, even when a nucleon participates in several collisions.} 
It would be tempting to speculate that these three hot spots effectively act as a single larger and elongated source, but this would lead to the opposite behavior to that of Fig.~\ref{fig:eigenvalues-Glauber}, namely a steeper spectrum for the MC Glauber model!
From that, we conclude that correlations do affect the spectrum of eigenvalues of the fluctuation modes in a non-trivial manner. 
However, one should note that in the present case the influence is rather small: comparing with Fig.~\ref{fig:eigenvalues1} or \ref{fig:eigenvalues2}, the difference in spectrum slope that we attribute to the correlations in the MC Glauber model are comparable to that which would be induced by a change in $\ssrc$ by 0.1~fm or less, although we did not attempt to quantitatively assess which value of $\ssrc$ would yield the same slope.

\section{Discussion}
\label{s:Discussion}

We used a semi-realistic model for the initial state of nuclear collisions, described as a superposition of independently distributed hot spots, to investigate the impact of different parameters --- the number of hot spots, their size, and the influence of hot-spot weights --- on the relative importance of modes characterizing the fluctuations of initial-state configurations about an average profile. 
Our main finding is that the greatest influence is that of the hot-spot size: 
a larger source width $\ssrc$ results in a steeper spectrum of the eigenvalues $\{\lambda_l\}$ characterizing the relative importance of the mode contributions to typical fluctuations, which means that fewer modes contribute significantly. 
In contrast, the number of hot spots $\Nsrc$ has little to no influence on the steepness of the spectra: 
$\Nsrc$ only affects (multiplicatively) the absolute value of the eigenstates, but this is paralleled by a similar increase of the average initial state $\bar{\Psi}(\vb{x})$. 
Letting the hot spots contribute with different weights to the initial states also has a small impact.

According to that view, the difference, mentioned at the end of Sec.~\ref{s:mode-decomposition}, in the slopes of the spectra of eigenvalues between the two initial-state models considered in Ref.~\cite{Borghini:2022iym} could reflect a difference in the size of the ``hot spots'' produced in the two models.
In the MC Glauber model, this source size is easily found: it is the smearing radius of 0.4~fm used for the energy density profile. 
In the saturation model of Ref.~\cite{Borghini:2022iym}, the energy density, given by Eq.~(21) of the article, is roughly proportional to the third power of the saturation scale ($Q_{s,A/B}$) in the colliding nuclei. 
Then the square of each $Q_s$ is proportional to the ``proton thickness function'', which introduces a typical length scale $\sqrt{B_G} = 2$~GeV$^{-1}\simeq 0.39$~fm. 
Since $Q_s$ is raised to the power $3/2$ in the expression of the energy density, the variance $B_G$ of the proton thickness function is effectively multiplied by $2/3$, which means that the typical size of the hot spots created in the model is approximately $\sqrt{2B_G/3} \simeq 0.32$~fm. 
This is smaller than the hot-spot size in the Glauber model, and can thus explain, to some extent, why the spectrum of eigenvalues $\{\lambda_l\}$ (or equivalently of weights $\{w_l\}$) is flatter in the saturation model than in the MC Glauber simulations at both values of the impact parameter studied in Ref.~\cite{Borghini:2022iym}.

Admittedly, the previous discussion ignores the effect of correlations in the initial state that are inherent in the models.
We have not studied such correlations here and reserve their detailed investigation for future work. 
The authors of Ref.~\cite{Blaizot:2014wba} concluded that the correlations they considered have a subleading impact on ``global'' observables like the geometry in the MC Glauber model.\footnote{In contrast, a close connection between correlations between nucleons in the initial state and multiparticle correlations in the final state was exhibited in Ref.~\cite{Giacalone:2023hwk}, arguing for a detailed experimental and phenomenological study of the latter to assess nuclear structure.}
Yet we found in Sec.~\ref{ss:results_IHSM_vs_MCGlauber} that the correlations present in the MC Glauber model lead to a sizable difference in the spectrum slope compared to IHSM simulations with the same average initial state and hot-spot size.
Intuition suggests that the correlations, being rather short-range, should affect the fluctuation modes with structure on shorter wavelengths, which are typically the modes with a smaller contribution to the initial states (with larger $l$ in our notations). 
At first sight, this expectation is not borne out by the results reported in Fig.~\ref{fig:Modes-IHSM(Glauber)}, since the whole spectrum is affected. 
One needs however to beware that the relative weights $w_l$ of the modes and $\bar{w}$ have to sum up to 1 by construction, so that a change for high-$l$ modes might entail a compensating shift in the small-$l$ modes --- unless the difference is absorbed by the average initial state. 
To test the impact of correlations more rigorously, one has to be able to turn them on and off in a well-controlled manner, for instance in a refined version of the hot-spot model used here.

In the present article, we mostly focused on the eigenvalues $\{\lambda_l\}$ associated with the modes $\{\Psi_l(\vb{x})\}$, but the latter were little discussed. 
As stated above, the findings across the various runs were quite similar, and thus Fig.~\ref{fig:Modes-example} represents the generic case, up to a few exceptions. 
First, the modes in the run with pointlike hot spots (not shown) are totally different. 
Instead of displaying regular features (rotational symmetry, dipole, quadrupole\dots) like the modes with extended sources, they are extremely chaotic, jumping wildly from positive to negative without any recognizable pattern from one cell of the discretized transverse plane to the next. 
This is consistent with the observation (Fig.~\ref{fig:eigenvalues1}) that the eigenvalues are extremely close to each other: since the modes are degenerate, modes with different symmetries mix, resulting in wildly looking modes.  
The only conspicuous trend is that the points where $\Psi_l(\vb{x})$ takes sizable values are close to the center for low $l$ and tend to extend further away as $l$ increases --- which is actually also the case for the modes of all runs, and can be seen in Fig.~\ref{fig:eigenvalues1} and ascribed to the higher probability to have fluctuations where the hot-spot-center density $f_1(\vb{x})$ is larger.

We already mentioned the second difference in behavior observed across the different runs, related to the integral of $\Psi_l(\vb{x})$ over the transverse plane, which for the sake of discussion we shall call the energy of the mode. 
This energy is vanishing (numerically: very small) for all modes in runs with a fixed number of hot spots $\Nsrc$ and fixed source weights $\varpi$, i.e.\ for all runs in which all initial configurations have exactly the same energy, which is then entirely contributed by the average initial state $\bar{\Psi}(\vb{x})$. 
Only in the runs with fluctuating $\Nsrc$ or $\varpi$ do a few modes have a finite energy. 
We recapitulate this here to further discuss the implications. 
Fixing the initial energy is what approximately happens when selecting events within a very narrow centrality class (as given by the multiplicity or the total energy at the end of a dynamical evolution), which is why we studied initial profiles with this constraint in this article. 
But we wish to emphasize that is not a fully innocuous constraint, since it actually induces small correlations between hot spots, even if the positions of the hot spots are independent (up to recentering). 
Indeed, intuition may suggest that a single hot spot might represent a fluctuation mode, in particular in the pointlike case.
This is however ruled out by the energy constraint, since a mode consisting of a single source would have a non-zero energy, so that at least a second hot spot with an opposite-sign contribution to the energy is required. 

To recover the intuition that single pointlike sources are the uncorrelated fluctuation modes $\{\Psi_l(\vb{x})\}$, a possibility is to consider configurations with unit weights and a fluctuating number of hot spots $\Nsrc$, such that the latter obeys Poisson statistics. 
In that case, the autocorrelation function~\eqref{eq:rho(x,y)_independent} simplifies to
\begin{equation}
\rho(\vb{x},\vb{y}) = \expval{\Nsrc}\!f_1(\vb{x})_{}\delta^{(2)}(\vb{x}-\vb{y}) =
\bar{\Psi}(\vb{x})_{}\delta^{(2)}(\vb{x}-\vb{y}),
\end{equation}
where we used Eq.~\eqref{eq:avg_state2} in the second identity. 
One immediately checks that $\delta^{(2)}(\vb{x}-\vb{x}_k)$ is an eigenvector of this function, with the eigenvalue $\lambda_l \equiv \bar{\Psi}(x_k)$. 
In that scenario, the Poissonian distribution of the hot-spot number means that the energy varies from event to event, lifting the corresponding constraint on the fluctuation modes.

\begin{acknowledgments}
We would like to thank Sören Schlichting for valuable discussions and Jean-Yves Ollitrault for clarifications on Ref.~\cite{Blaizot:2014wba}.
N.~B. and H.~R. acknowledge support by the Deutsche Forschungsgemeinschaft (DFG, German Research Foundation) through the CRC-TR 211 'Strong-interaction matter under extreme conditions' - project number 315477589 - TRR 211.
H.~R. was supported in part by the National Science Foundation (NSF) within the framework of the JETSCAPE collaboration (OAC-2004571) and by the DOE (DE-SC0024232).
Numerical simulations presented in this work were performed at the Paderborn Center for Parallel Computing (PC$^2$) and we gratefully acknowledge their support.
\end{acknowledgments}

\appendix

\section{Statistics of the expansion coefficients}
\label{s:c_l}

In this Appendix, we present a few illustrative results from one of our sets of simulations --- the ``reference'' run with rotationally symmetric hot-spot density, $\Nsrc=50$ sources with width $\ssrc=0.3$~fm and unit weights --- pertaining to the coefficients $\{c_l\}$ in the expansion~\eqref{eq:evt-fluct_vs_modes} of fluctuations on a basis of modes.

\begin{figure}[b!]
\includegraphics[width=\linewidth]{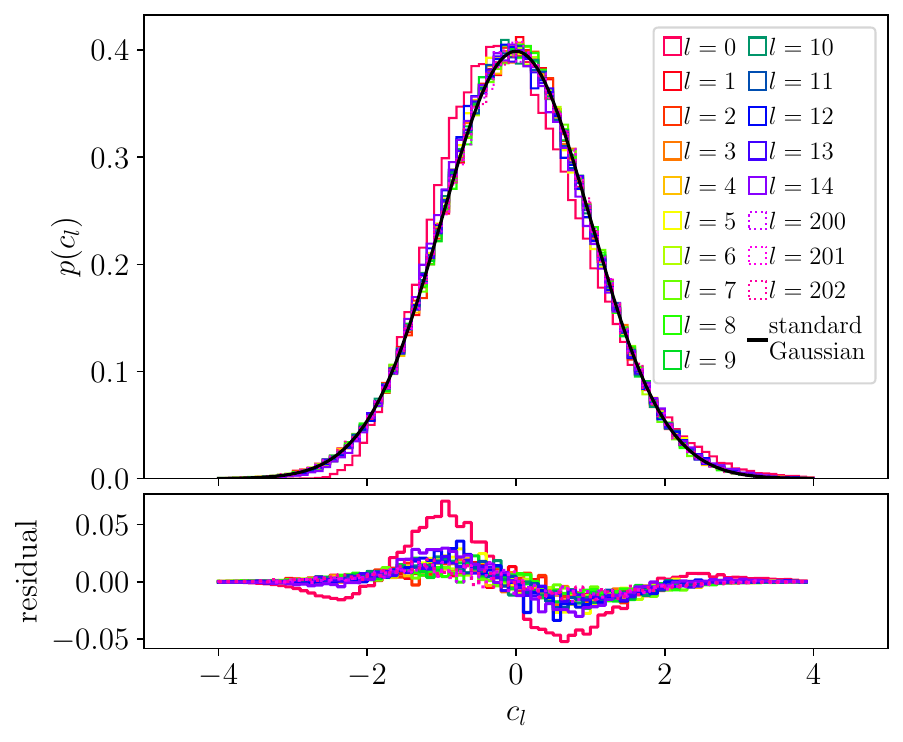}\vspace{-2.5mm}
\caption{Frequency histograms of the expansion coefficients $c_l$ for a few modes of run IHSM(4,4)$_{50}^{0.3}$, compared with a standard Gaussian distribution.}
\label{fig:cl-dist_R3}
\end{figure}

For that purpose, we randomly selected $2^{17}$ configurations among the $2^{21}$ of the run, and decomposed their fluctuation part $\delta\Phi^{(i)}$ over the basis of eigenstates (determined from the whole run), to obtain the expansion coefficients $\{c_l^{(i)}\}$. 
In Fig.~\ref{fig:cl-dist_R3} we show histograms of the observed $c_l$ distributions of a few fluctuation modes, namely the most important ones (small $l$), and three higher modes.
By construction, the average value of a given $c_l$ should be zero and its variance unity. 
But one can see that the probability distribution is to a good approximation Gaussian, except for a small skewness, which becomes marked for the mode $l=0$ --- for which we have no clear explanation. 

\begin{figure}[t!]
\includegraphics[width=\linewidth]{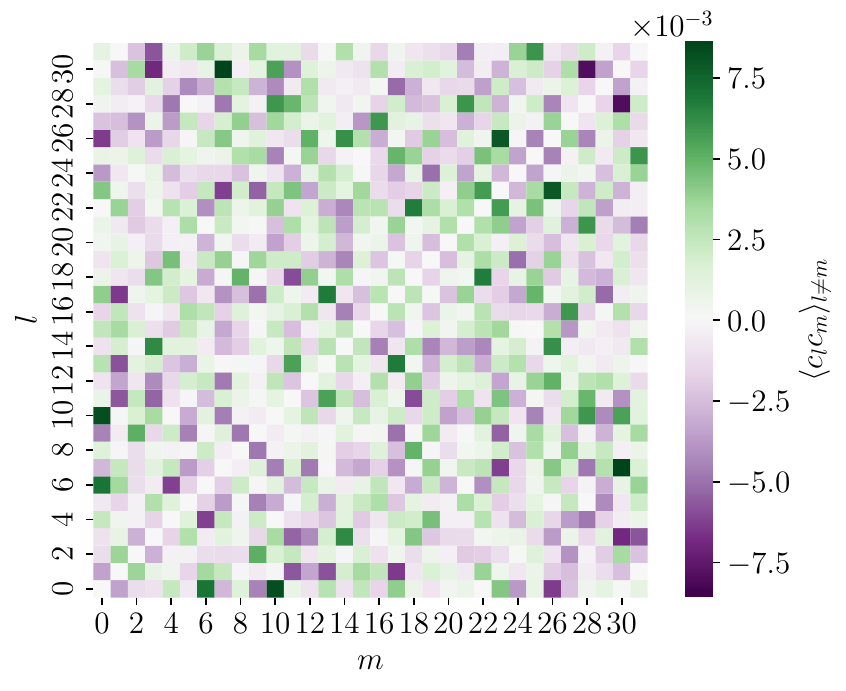}\vspace{-2.5mm}
\caption{Cross-correlation $\expval{c_l c_m}$ of the expansion coefficients over two different modes of run IHSM(4,4)$_{50}^{0.3}$. The terms on the diagonal are approximately equal to 1 by construction and not shown.}
\label{fig:<cl_cm>_R3}
\end{figure}

In Fig.~\ref{fig:<cl_cm>_R3} we display the cross-correlation $\expval{c_l c_m}_{l\neq m}$ of the expansion coefficients along different fluctuation modes.\footnote{Note that these two-point averages do not coincide with the respective covariances, since numerically the average value of the expansion coefficients $c_l$ is not exactly zero for the sample of initial configurations used for the results of this Appendix.}
The values of those averages are systematically smaller than $10^{-2}$, much smaller than the averages $\expval{c_l^2}\simeq 1$, which shows that we indeed obtained uncorrelated modes, Eq.~\eqref{eq:mean_and_variance_cl2}.

\section{Fluctuation modes for the run IHSM(Glauber)$_{250}^{0.4}$}
\label{s:modes_ISHM(Glauber)}

In this Appendix we present the first 60 orthonormal eigenvectors corresponding to the fluctuation modes for the run IHSM(Glauber)$_{250}^{0.4}$ described in Sec.~\ref{ss:results_IHSM_vs_MCGlauber}.
\begin{figure*}
	\includegraphics[height=0.975\textheight]{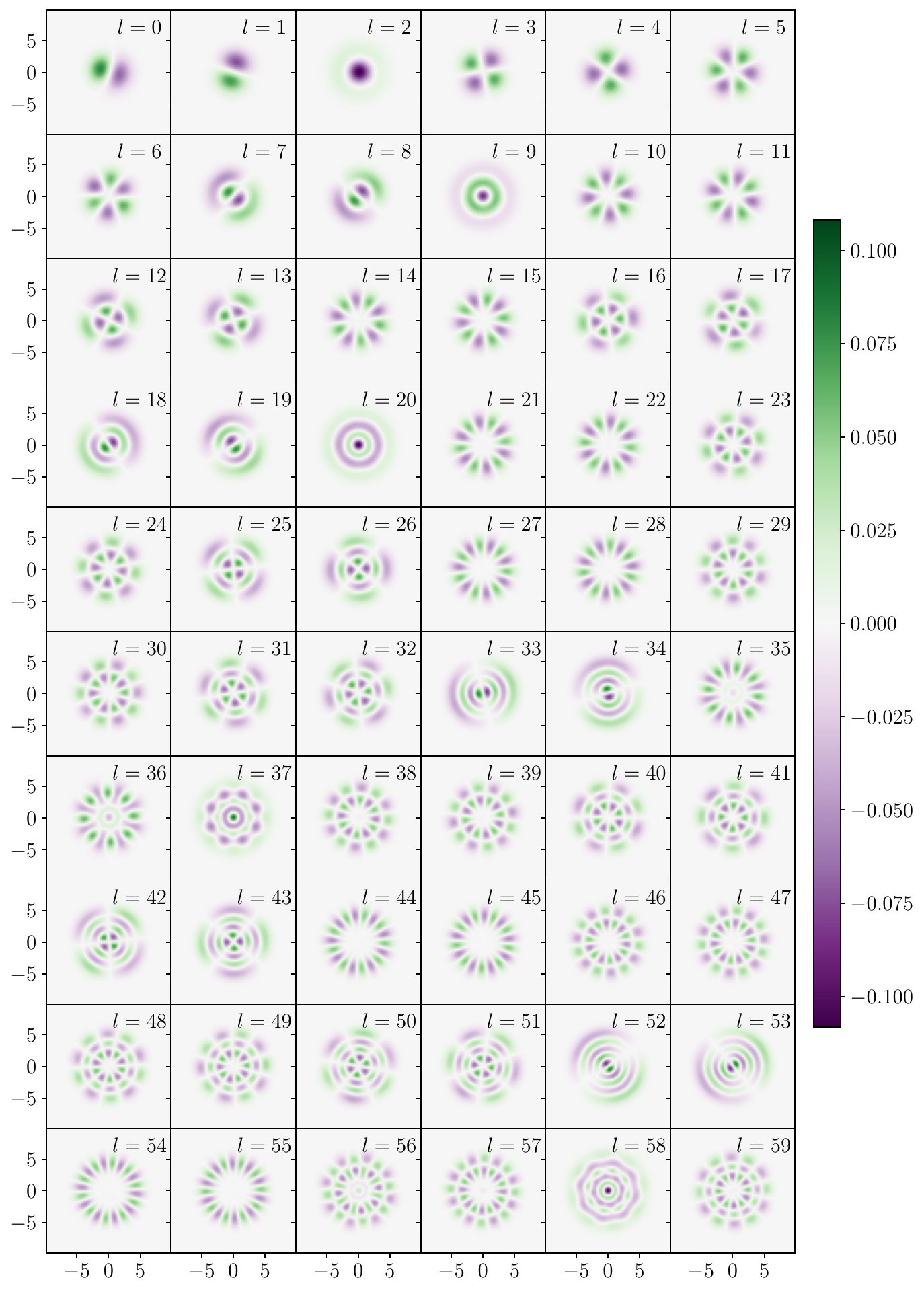}
	\vspace{-3mm}
	\caption{Density plots of the first 60 orthonormal eigenvectors for the run IHSM(Glauber)$_{250}^{0.4}$. Both axes are in units of fm.}
	\label{fig:Modes-IHSM(Glauber)}
\end{figure*}
These should be compared to those from MC Glauber simulations at $b=0$ displayed in Fig.~30 of Ref.~\cite{Borghini:2022iym}.

\clearpage

\begin{widetext}
	
	\section*{Supplemental material}
	
	\subsection*{Average initial states}\vspace{-2mm}
	
	\begin{figure}[!h]
		\includegraphics[width=0.45\linewidth]{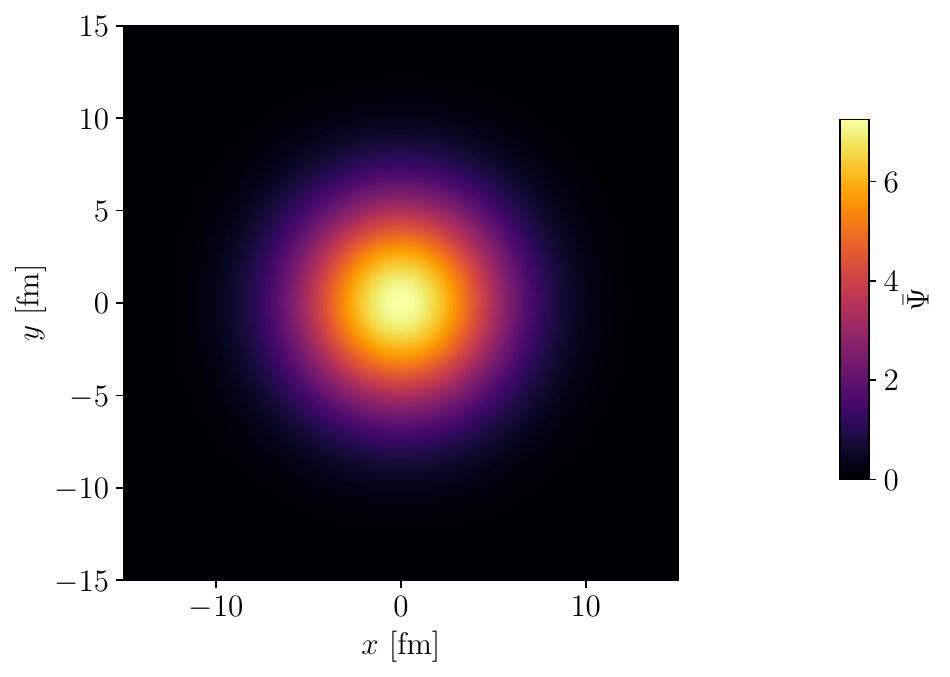}\hspace{10mm}
		\includegraphics[width=0.46\linewidth]{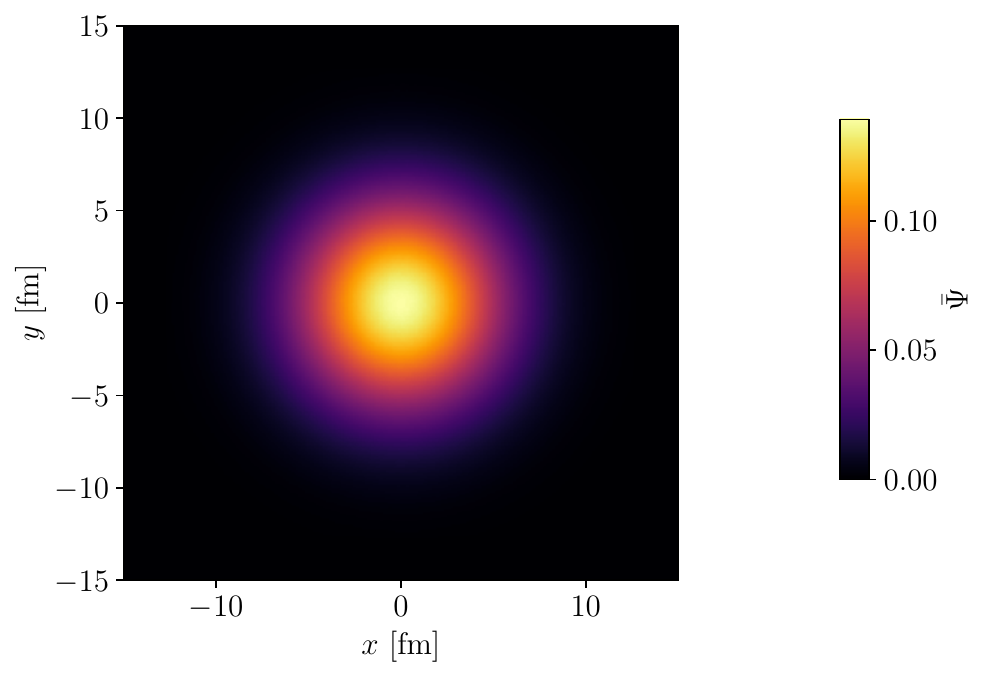}\vspace{-3mm}
		\caption{Density profile of the average initial state $\bar{\Psi}(\vb{x})$ for the runs IHSM(4,4)$_{750}^{0.7}$ (left) and IHSM(4,4)$_{250}^{0.0}$ (right).}\vspace{-4mm}
		\label{fig:av-state-IHSM(4,4)_750^0.7&(4,4)_250^0}
	\end{figure}
	\begin{figure}[!h]
		\includegraphics[width=0.45\linewidth]{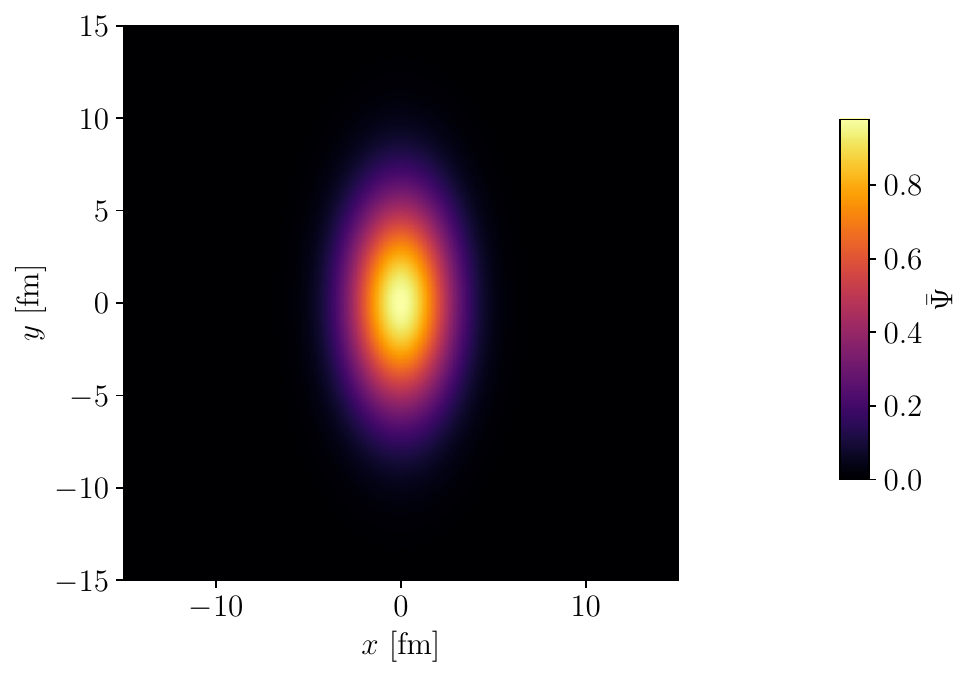}\vspace{-3mm}
		\caption{Density profile of the average initial state $\bar{\Psi}(\vb{x})$ for the runs IHSM(2,4)$_{50}^{0.3}$.}
		\label{fig:av-state-IHSM(2,4)_50^0.3}\vspace{-2mm}
	\end{figure}
	
	Here, we display the density profile of the average initial state $\bar{\Psi}(\vb{x})$ for three sample runs, namely two with a rotationally symmetric distribution of hot spots (Figs.~\ref{fig:av-state-IHSM(4,4)_750^0.7&(4,4)_250^0}) and one with an elongated distribution (Fig.~\ref{fig:av-state-IHSM(2,4)_50^0.3}). 
	
	All runs with the same underlying hot-spot-density are almost indistinguishable by eye, as illustrated by the two examples of Fig.~\ref{fig:av-state-IHSM(4,4)_750^0.7&(4,4)_250^0} (see also Fig.~\ref{fig:av-state}): 
	On the left is the result for the run with configurations consisting of $\Nsrc=750$ sources with width $\ssrc=0.7$~fm, and on the right is the result for the run with configurations with 250 pointlike sources, yet one can hardly recognize that the former profile is slightly broader than the latter, as indicated by the mean square widths reported in Sec.~\ref{ss:results_av-state}. 
	
	In turn, the average initial states of all four runs with an elongated hot-spot density look almost the same as that shown in Fig.~\ref{fig:av-state-IHSM(2,4)_50^0.3}.

	\subsection*{Fluctuation modes}
	
	In Figs.~\ref{fig:Modes-IHSM(4,4)_250^0.3}--\ref{fig:Modes-fluctuating_ssrc} we display the first sixty normalized eigenvectors $\Psi_l/\sqrt{\lambda_l}$ associated with the fluctuation modes for the runs listed in Table~\ref{tab:sim_parameters}. 
	In the runs with finite-size sources and rotationally symmetric distribution of hot spots, the modes are qualitatively similar to those of Fig.~\ref{fig:Modes-example} of the article, even when one of the simulation parameters is fluctuating (Figs.~\ref{fig:Modes-fluctuating_N}--\ref{fig:Modes-fluctuating_ssrc}).
	
	Then the modes for runs with an elongated distribution of hot spots (Figs.~\ref{fig:Modes-IHSM(2,4)_50^0.3}--\ref{fig:Modes-IHSM(2,4)_250^0.7}) also show some recognizable elements of symmetry, although the $x$- and $y$-directions now clearly play a special role. 
	
	Eventually, the modes for the run with configurations consisting of $\Nsrc=50$ pointlike hot spots (Fig.~\ref{fig:Modes-IHSM(4,4)_250^0.0}) are quite different, which is due to their approximate degeneracy, as discussed in the text of the article.  
	
	\begin{figure*}
		\includegraphics[height=0.975\textheight]{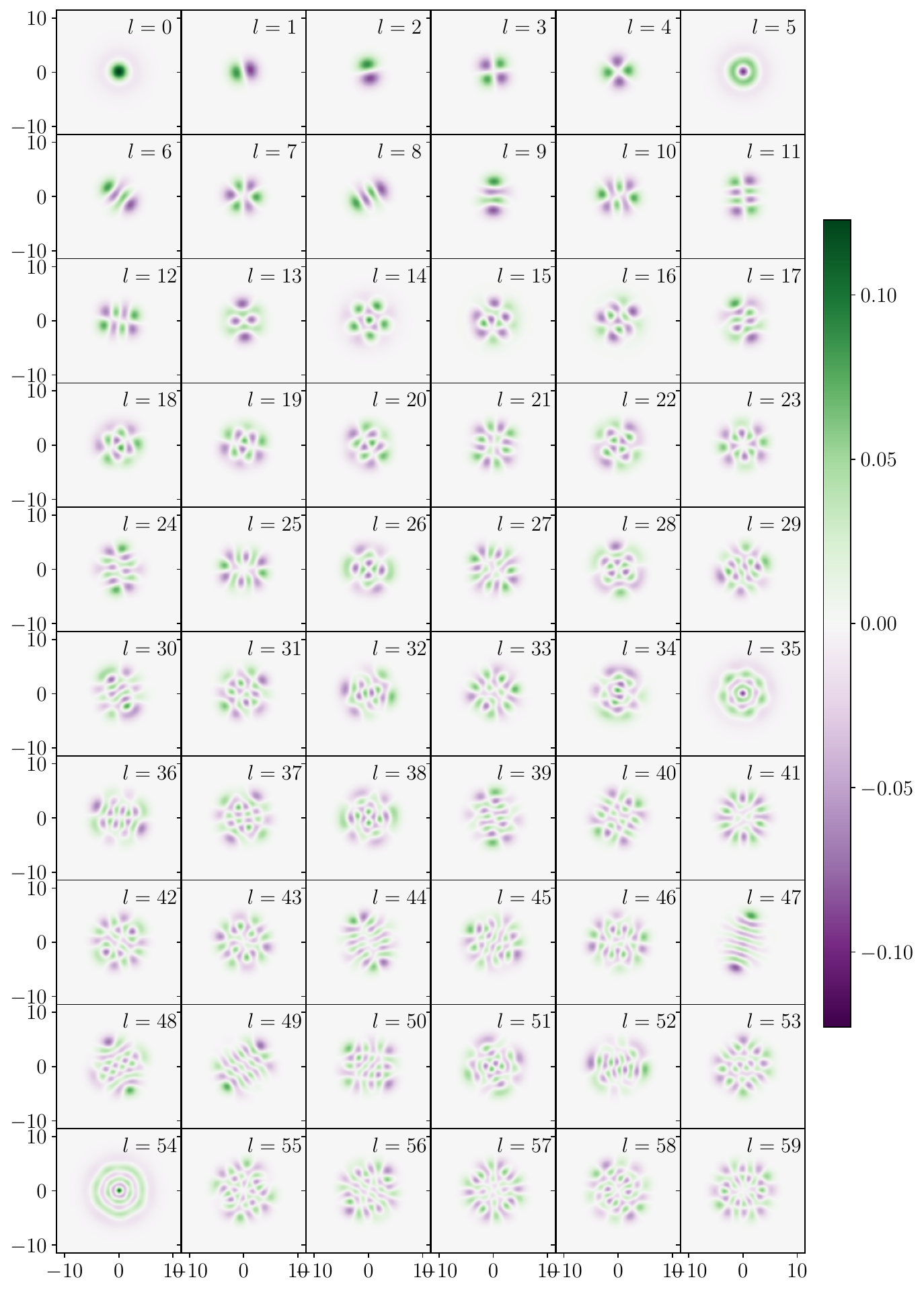}
		\vspace{-3mm}
		\caption{Density plots of the first 60 orthonormal eigenvectors for the run IHSM(4,4)$_{250}^{0.3}$. Both axes are in units of fm.}
		\label{fig:Modes-IHSM(4,4)_250^0.3}
	\end{figure*}
	
	\begin{figure*}
		\includegraphics[height=0.975\textheight]{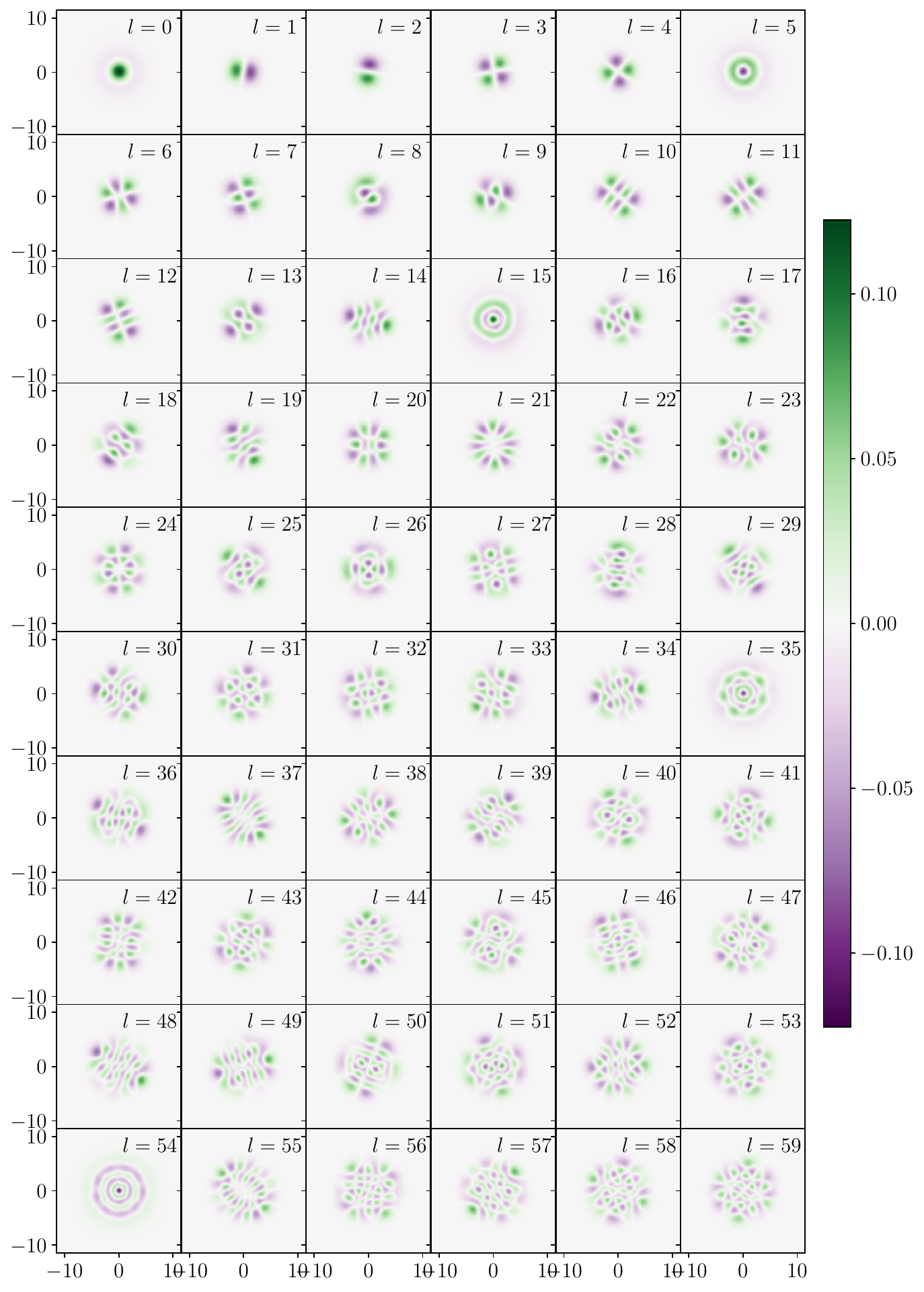}
		\vspace{-3mm}
		\caption{Density plots of the first 60 orthonormal eigenvectors for the run IHSM(4,4)$_{750}^{0.3}$. Both axes are in units of fm.}
		\label{fig:Modes-IHSM(4,4)_750^0.3}
	\end{figure*}
	
	\begin{figure*}
		\includegraphics[height=0.975\textheight]{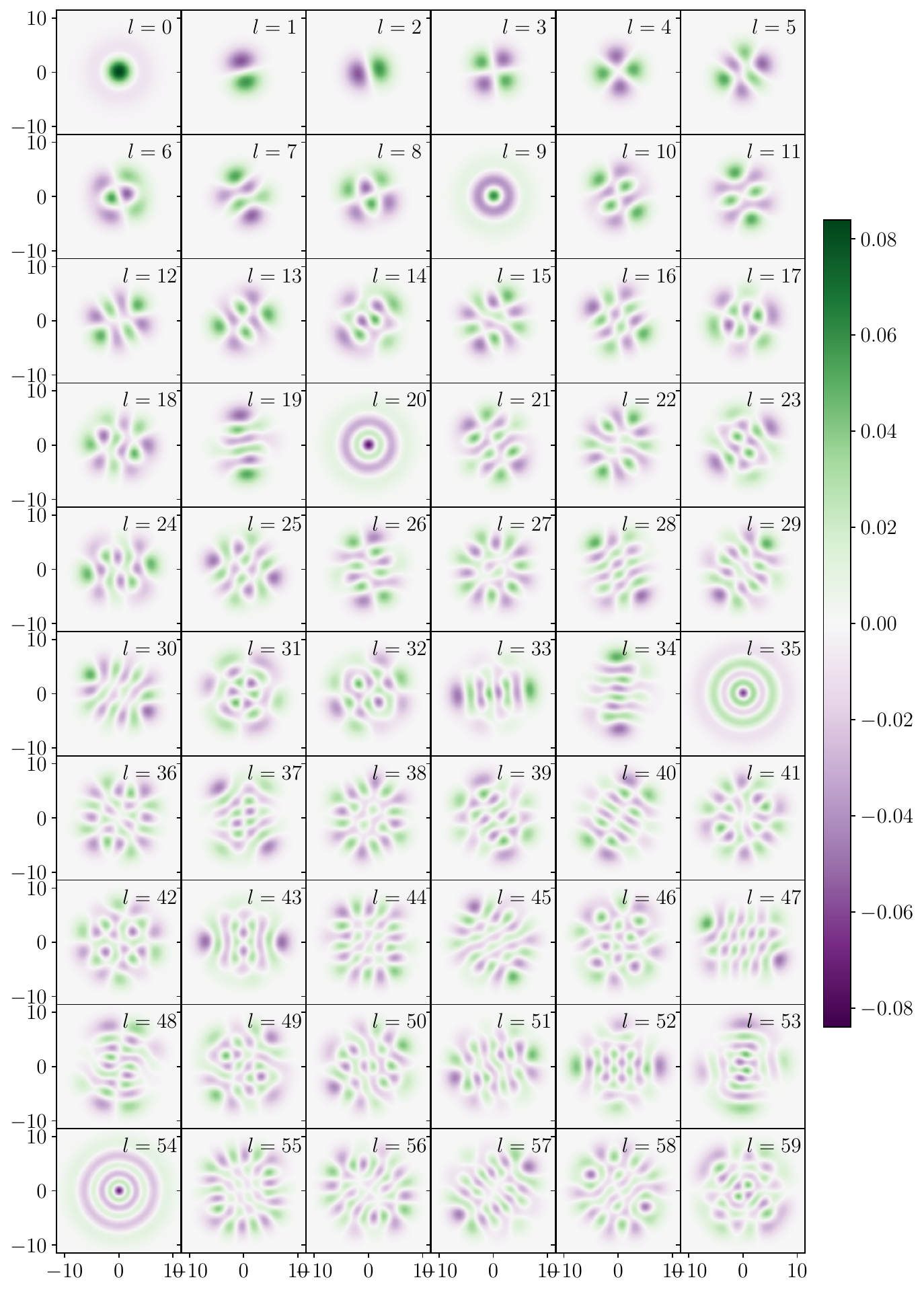}
		\vspace{-3mm}
		\caption{Density plots of the first 60 orthonormal eigenvectors for the run IHSM(4,4)$_{50}^{0.7}$. Both axes are in units of fm.}
		\label{fig:Modes-IHSM(4,4)_50^0.7}
	\end{figure*}
	
	\begin{figure*}
		\includegraphics[height=0.975\textheight]{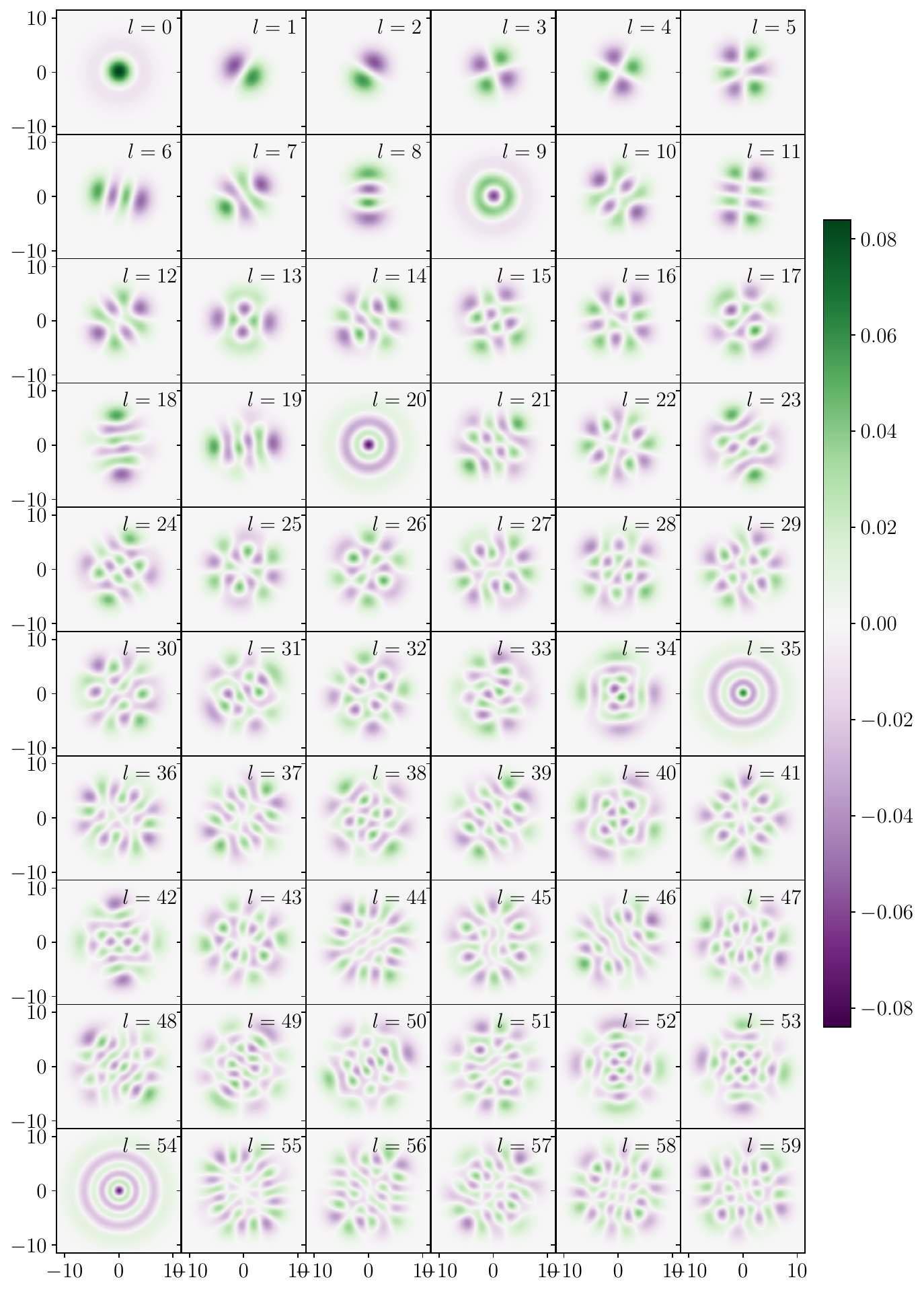}
		\vspace{-3mm}
		\caption{Density plots of the first 60 orthonormal eigenvectors for the run IHSM(4,4)$_{250}^{0.7}$. Both axes are in units of fm.}
		\label{fig:Modes-IHSM(4,4)_250^0.7}
	\end{figure*}
	
	\begin{figure*}
		\includegraphics[height=0.975\textheight]{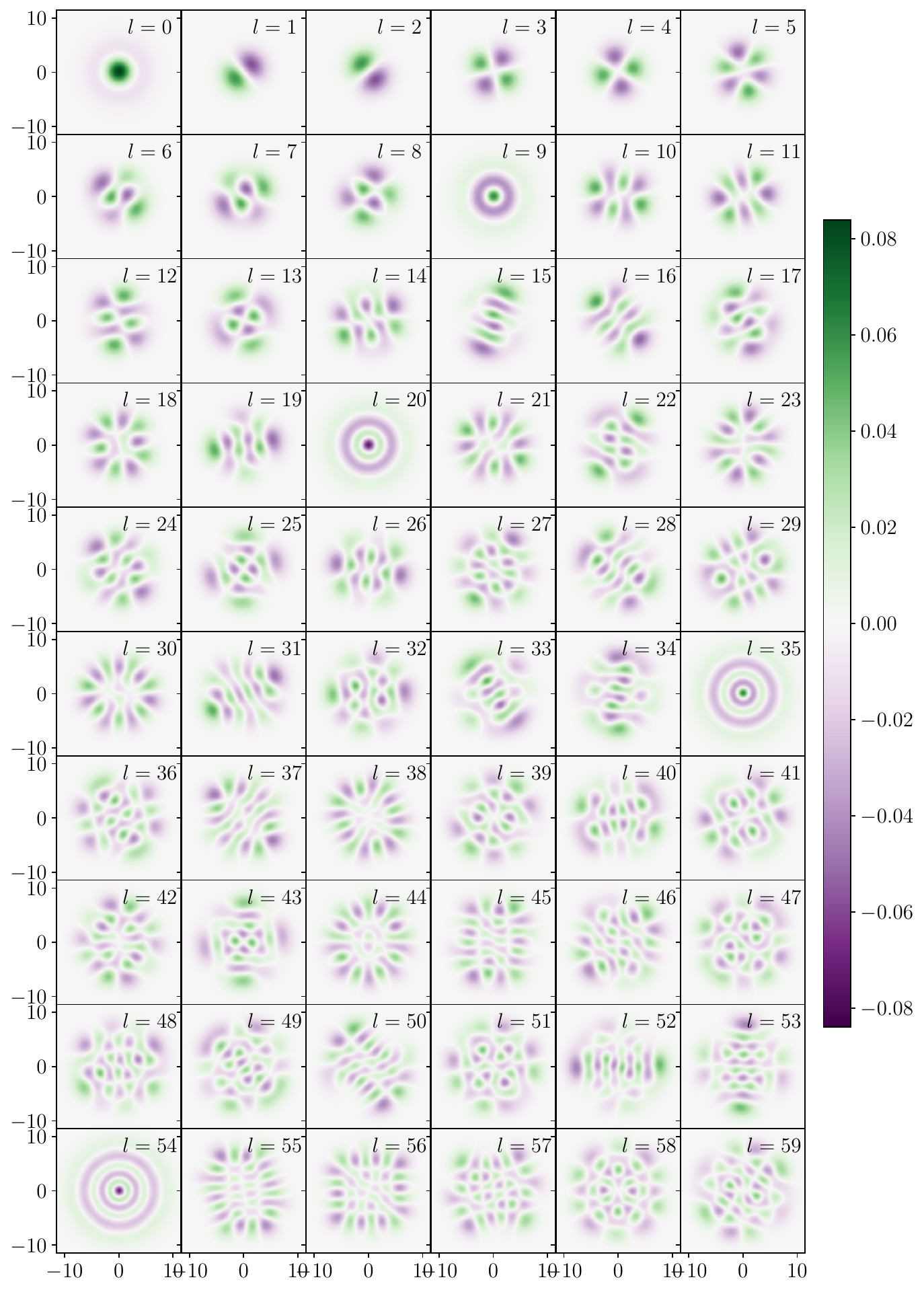}
		\vspace{-3mm}
		\caption{Density plots of the first 60 orthonormal eigenvectors for the run IHSM(4,4)$_{750}^{0.7}$. Both axes are in units of fm.}
		\label{fig:Modes-IHSM(4,4)_750^0.7}
	\end{figure*}
	
	\begin{figure*}
		\includegraphics[height=0.975\textheight]{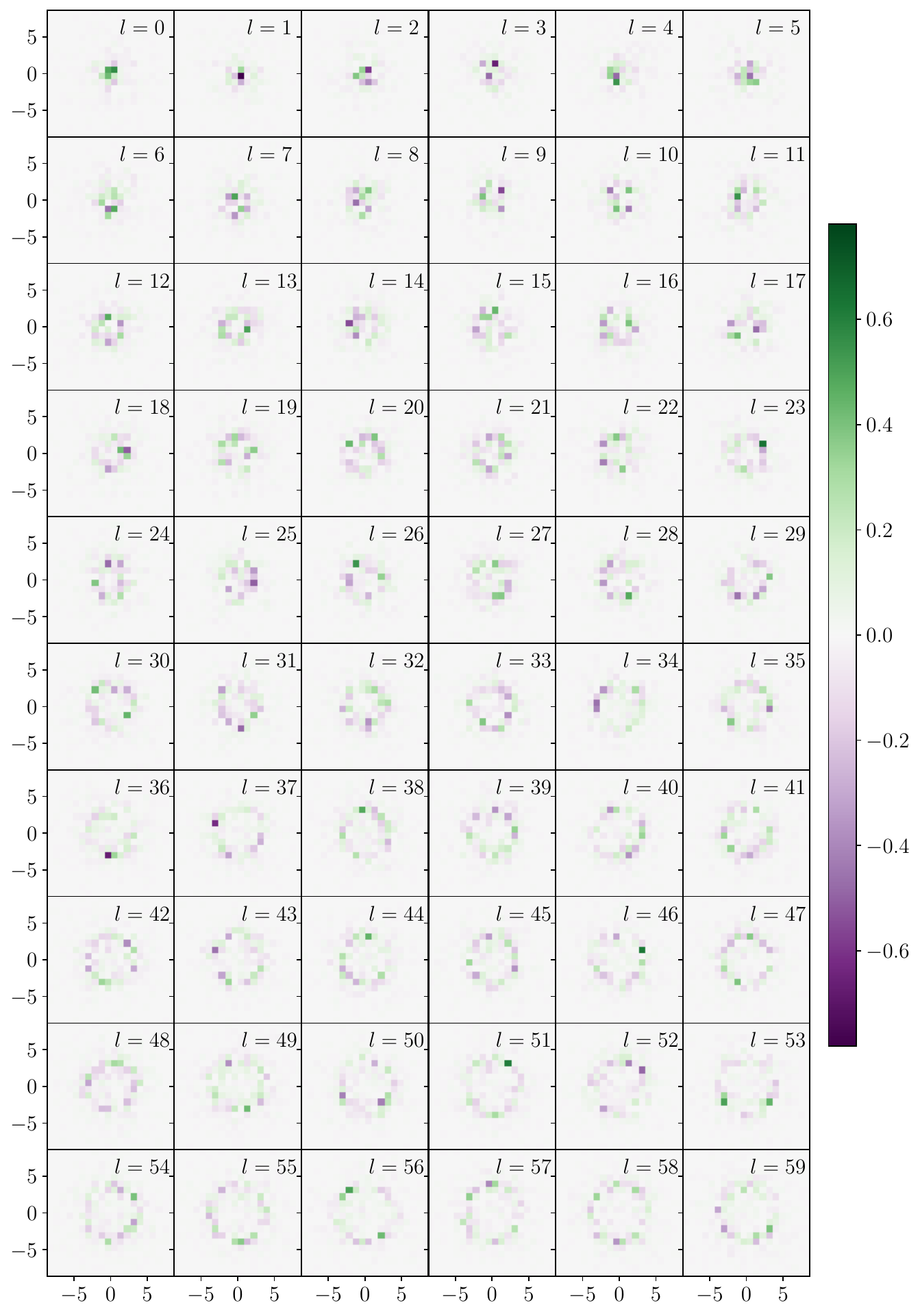}
		\vspace{-3mm}
		\caption{Density plots of the first 60 orthonormal eigenvectors for the run IHSM(4,4)$_{250}^{0.0}$. Both axes are in units of fm.}
		\label{fig:Modes-IHSM(4,4)_250^0.0}
	\end{figure*}
	
	\begin{figure*}
		\includegraphics[height=0.975\textheight]{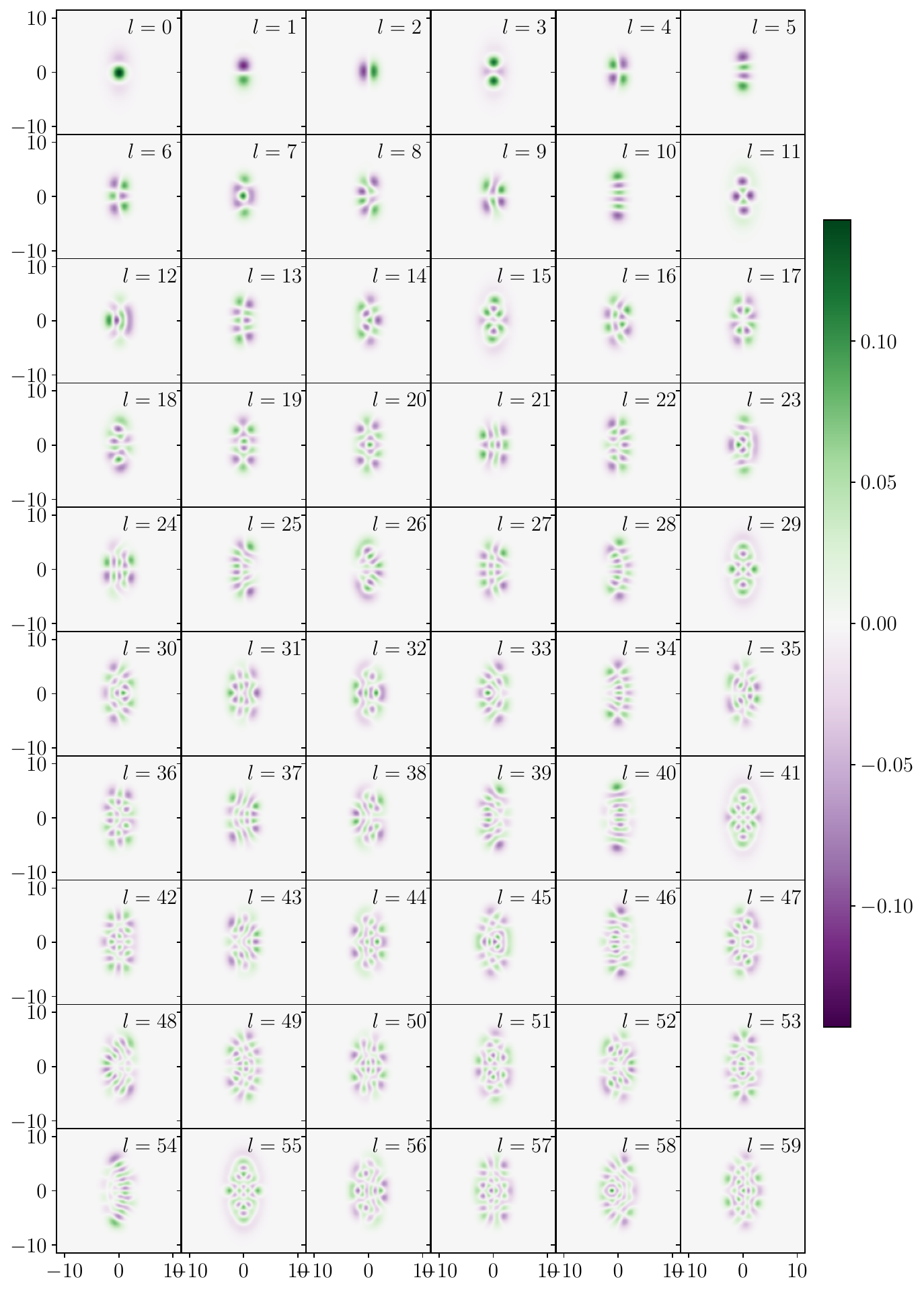}
		\vspace{-3mm}
		\caption{Density plots of the first 60 orthonormal eigenvectors for the run IHSM(2,4)$_{50}^{0.3}$. Both axes are in units of fm.}
		\label{fig:Modes-IHSM(2,4)_50^0.3}
	\end{figure*}
	
	\begin{figure*}
		\includegraphics[height=0.975\textheight]{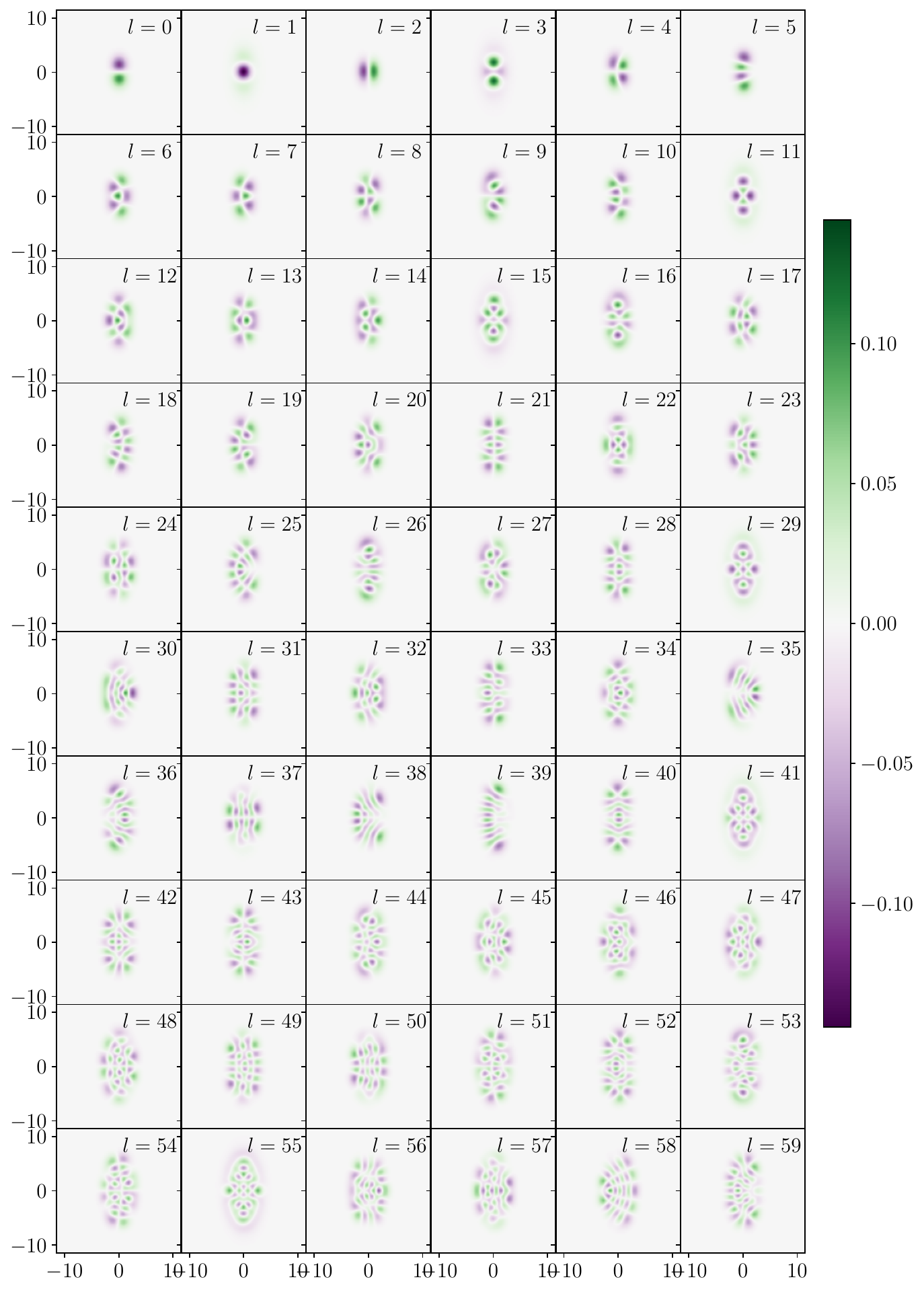}
		\vspace{-3mm}
		\caption{Density plots of the first 60 orthonormal eigenvectors for the run IHSM(2,4)$_{250}^{0.3}$. Both axes are in units of fm.}
		\label{fig:Modes-IHSM(2,4)_250^0.3}
	\end{figure*}
	
	\begin{figure*}
		\includegraphics[height=0.975\textheight]{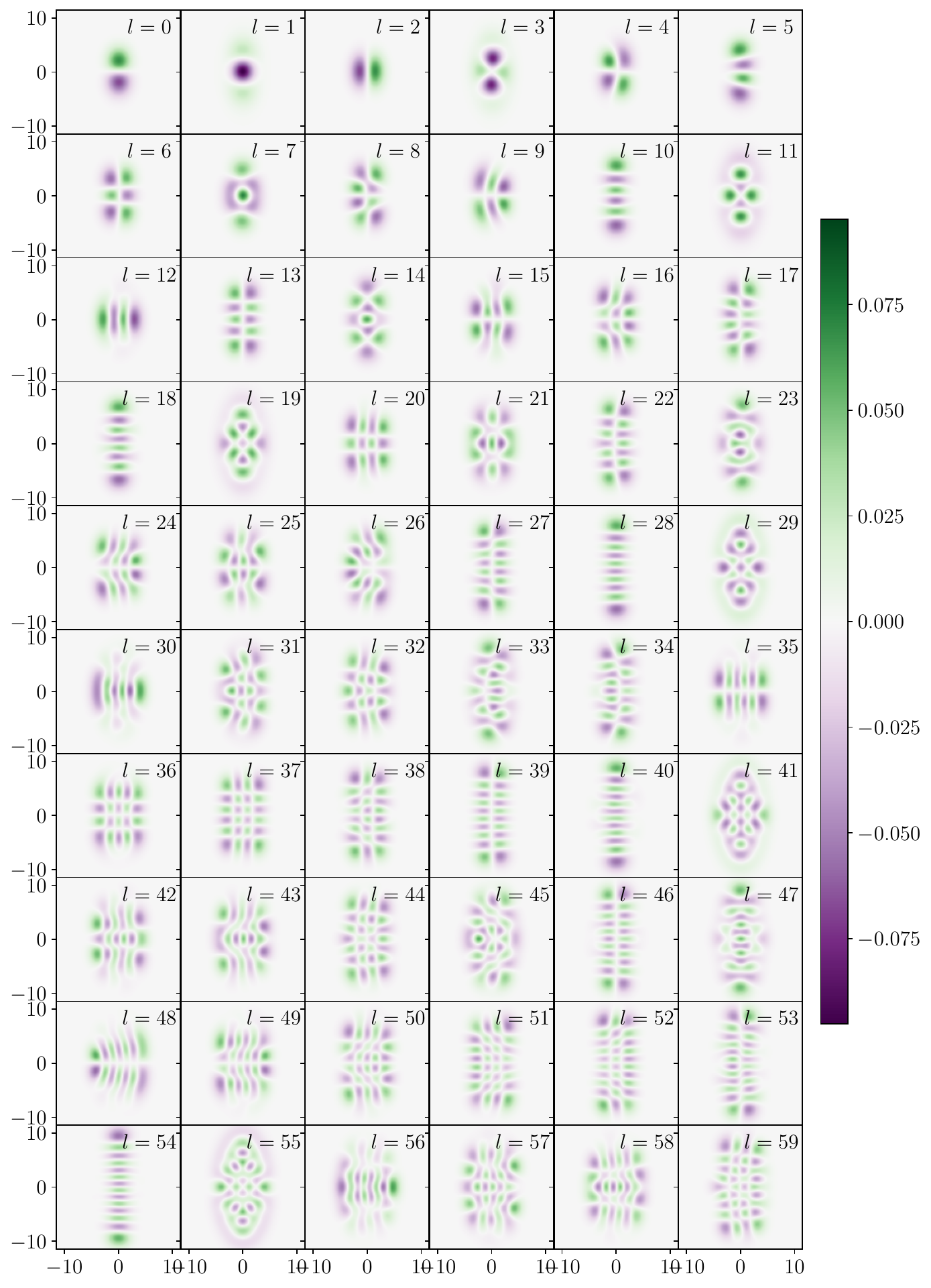}
		\vspace{-3mm}
		\caption{Density plots of the first 60 orthonormal eigenvectors for the run IHSM(2,4)$_{50}^{0.7}$. Both axes are in units of fm.}
		\label{fig:Modes-IHSM(2,4)_50^0.7}
	\end{figure*}
	
	\begin{figure*}
		\includegraphics[height=0.975\textheight]{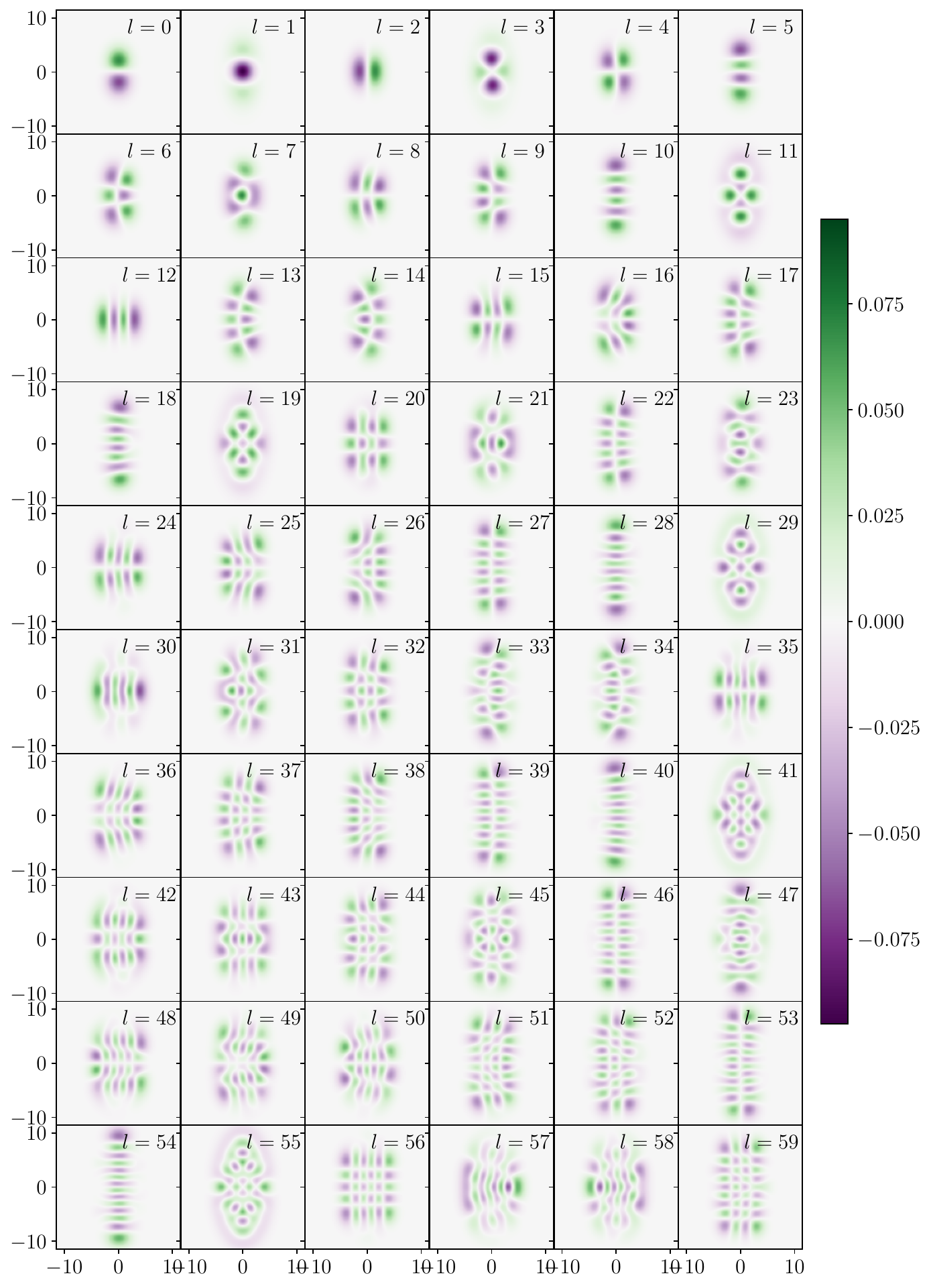}
		\vspace{-3mm}
		\caption{Density plots of the first 60 orthonormal eigenvectors for the run IHSM(2,4)$_{250}^{0.7}$. Both axes are in units of fm.}
		\label{fig:Modes-IHSM(2,4)_250^0.7}
	\end{figure*}
	
	\begin{figure*}
		\includegraphics[height=0.975\textheight]{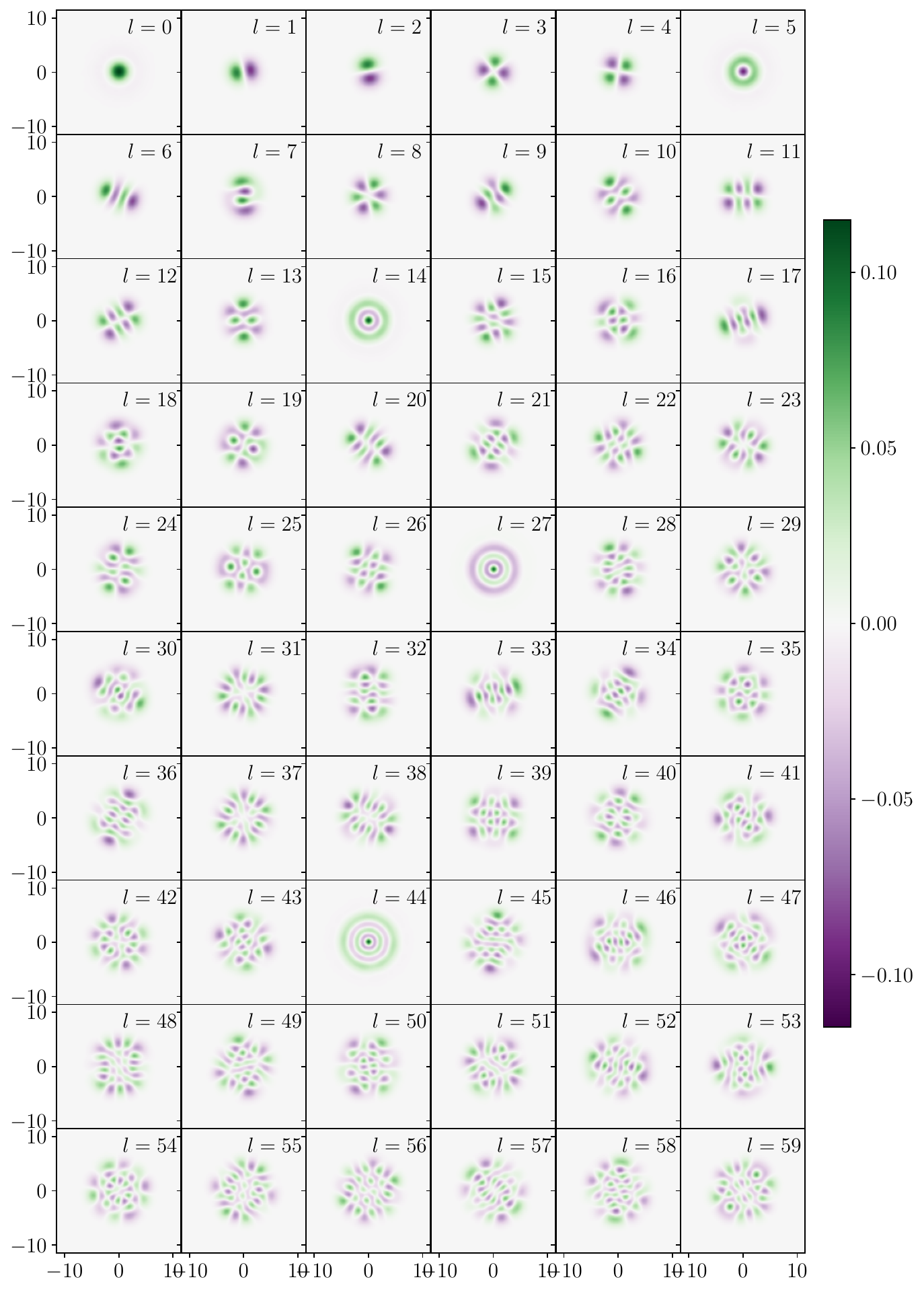}
		\vspace{-3mm}
		\caption{Density plots of the first 60 orthonormal eigenvectors for the run $\Nsrc=50\pm 10$. Both axes are in units of fm.}
		\label{fig:Modes-fluctuating_N}
	\end{figure*}
	
	\begin{figure*}
		\includegraphics[height=0.975\textheight]{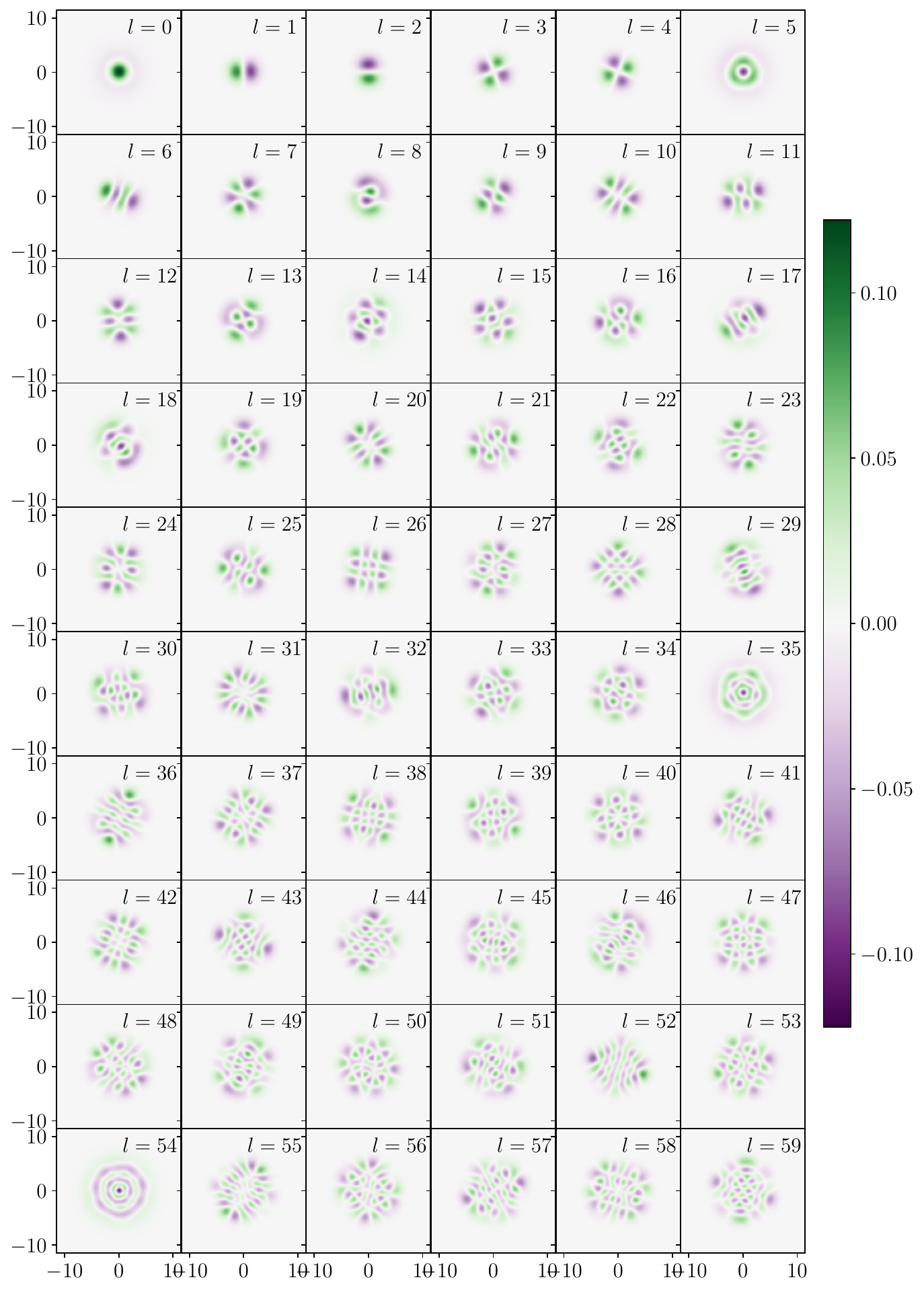}
		\vspace{-3mm}
		\caption{Density plots of the first 60 orthonormal eigenvectors for the run $\varpi=1\pm 0.3$. Both axes are in units of fm.}
		\label{fig:Modes-fluctuating_w}
	\end{figure*}
	
	\begin{figure*}
		\includegraphics[height=0.975\textheight]{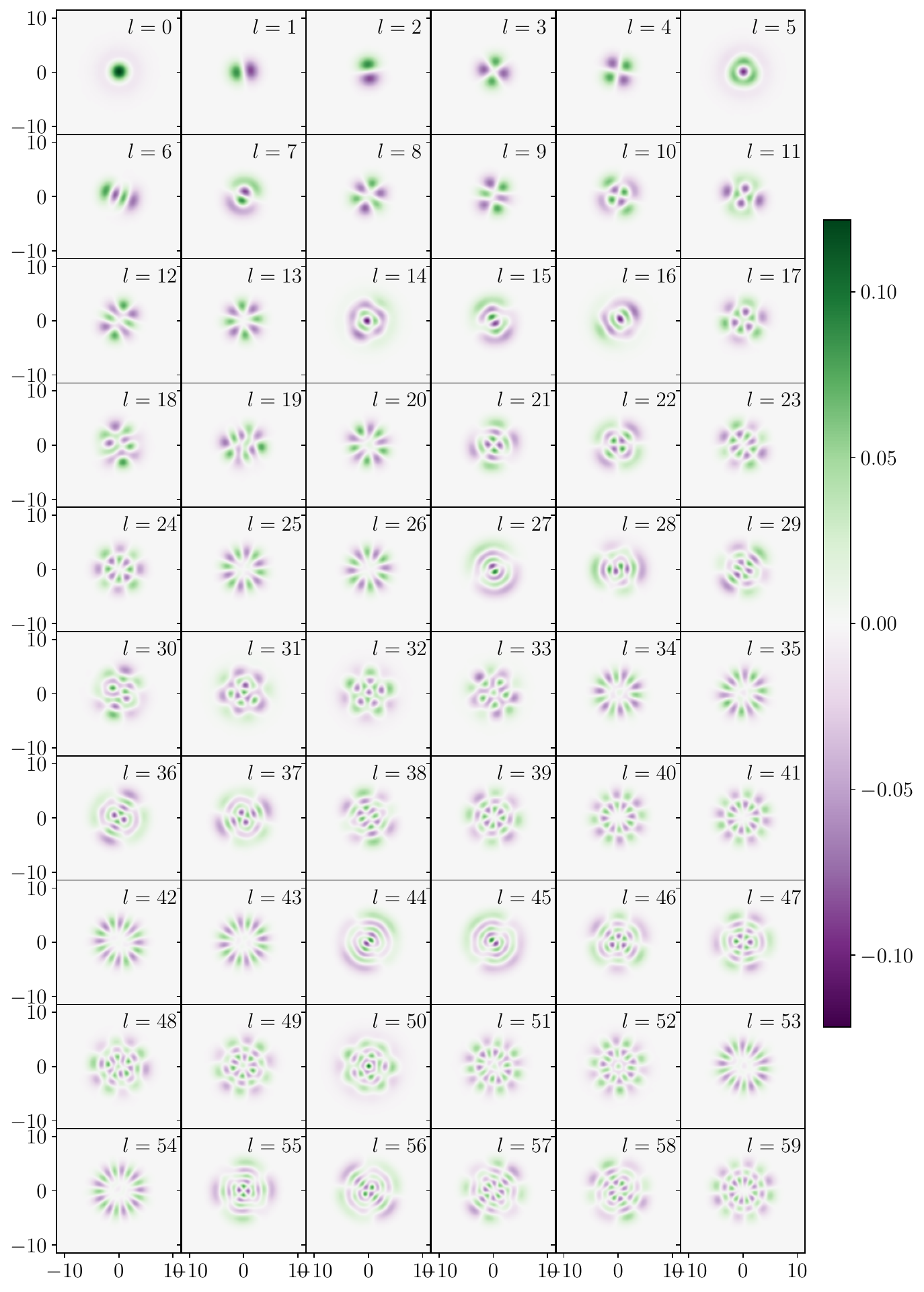}
		\vspace{-3mm}
		\caption{Density plots of the first 60 orthonormal eigenvectors for the run $\ssrc=0.3\pm 0.17$~fm. Both axes are in units of fm.}
		\label{fig:Modes-fluctuating_ssrc}
	\end{figure*}

	\subsection*{Eigenvalues}

	Eventually, we gather in Fig.~\ref{fig:eigenvalues2vs4} the eigenvalues $\{\lambda_l\}$ of the fluctuation modes for the runs with initial configurations with $\Nsrc=50$ or 250 and $\ssrc=0.3$ or 0.7~fm for both rotationally symmetric (already shown in Fig.~\ref{fig:eigenvalues1}, top panel) and elongated (Fig.~\ref{fig:eigenvalues2}, top panel) hot-spot distributions.
	This allows one to visualize at once the influence on the $\{\lambda_l\}$ of the underlying geometry. 
	\begin{figure}[!h]
		\includegraphics[width=.5\linewidth]{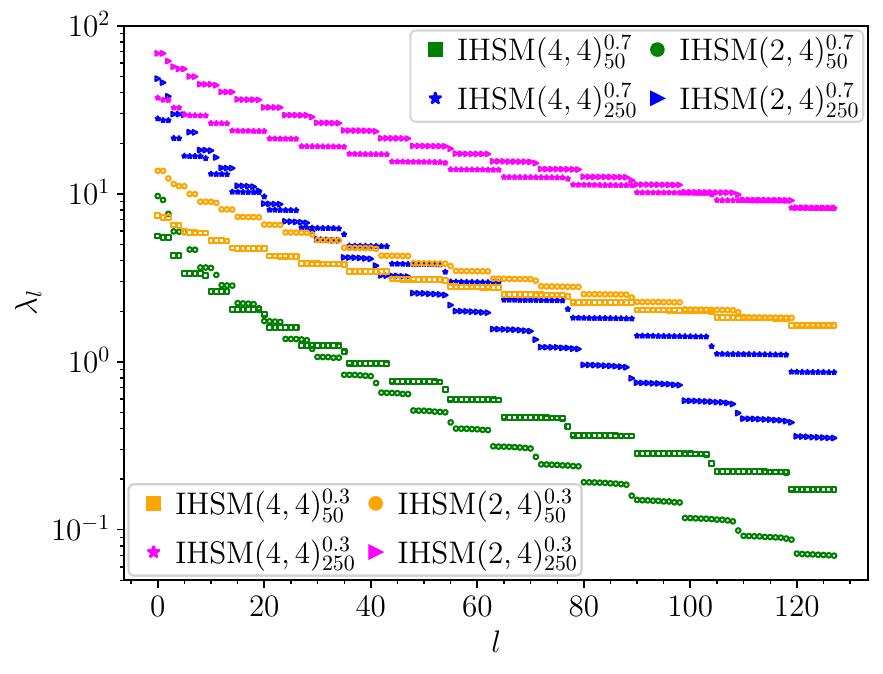}\vspace{-2mm}
		\caption{Eigenvalues $\lambda_l$ of the fluctuation modes in simulations with a rotationally symmetric or elongated hot-spot distributions.}
		\label{fig:eigenvalues2vs4}
	\end{figure}
	
	As already mentioned in the article, the ``quasi-degenerate plateaus'' involve fewer modes when the geometry is no longer rotationally symmetric, and the plateaus are less flat, consistent with the degeneracy lifting due to symmetry breaking. 
	
	Another trend is that, at fixed $\Nsrc$ and $\ssrc$, the $\lambda_l$-spectrum is steeper in the run with an elongated hot-spot distribution. 
	This behavior can be ascribed to the fact that the sources overlap more in that case than in the rotationally symmetric scenario. 
	Indeed, due to the smaller area over which the hot spots are distributed, their number density is higher. 
	Effectively, this has the same consequence as increasing the hot-spot size, namely, this suppresses fluctuation modes with smaller wavelengths.

\end{widetext}

\end{document}